\documentclass[11pt]{article}
\usepackage{graphicx}

\begin{document}

\def\xslash#1{{\rlap{$#1$}/}}
\def \p {\partial}
\def \dd {\psi_{u\bar dg}}
\def \ddp {\psi_{u\bar dgg}}
\def \pq {\psi_{u\bar d\bar uu}}
\def \jpsi {J/\psi}
\def \psip {\psi^\prime}
\def \to {\rightarrow}
\def\bfsig{\mbox{\boldmath$\sigma$}}
\def\DT{\mbox{\boldmath$\Delta_T $}}
\def\xit{\mbox{\boldmath$\xi_\perp $}}
\def \jpsi {J/\psi}
\def\bfej{\mbox{\boldmath$\varepsilon$}}
\def \t {\tilde}
\def\epn {\varepsilon}
\def \up {\uparrow}
\def \dn {\downarrow}
\def \da {\dagger}
\def \pn3 {\phi_{u\bar d g}}

\def \p4n {\phi_{u\bar d gg}}

\def \bx {\bar x}
\def \by {\bar y}


\begin{center}
{\Large\bf  Collinear Factorization for Single Transverse-Spin Asymmetry in Drell-Yan Processes }
\par\vskip20pt
J.P. Ma$^{1,2}$, H.Z. Sang$^{3}$ and S.J. Zhu$^{1}$    \\
{\small {\it
$^1$ Institute of Theoretical Physics, Academia Sinica,
P.O. Box 2735,
Beijing 100190, China\\
$^2$ Center for High-Energy Physics, Peking University, Beijing 100871, China  \\
$^3$ Institute of Modern Physics,
School of Science,
East China University of Science and Technology,
130 Meilong Road, Shanghai 200237, P.R. China
}} \\
\end{center}
\vskip 1cm
\begin{abstract}
We study the scattering of a single parton state with a multi-parton state to derive the complete results of perturbative coefficient functions at leading order, which appear in the collinear factorization for Single transverse-Spin Asymmetry(SSA) in Drell-Yan processes with a transversely polarized hadron in the initial state. 
We find that the factorization formula of SSA contains
hard-pole-, soft-quark-pole- and soft-gluon-pole contributions.
It is interesting to note that the leading order perturbative coefficient functions of soft-quark-pole- and soft-gluon-pole contributions are extracted from parton scattering amplitudes at one-loop, while the functions of hard-pole 
contributions are extracted from the tree level amplitudes at tree-level. 
Our method to derive the factorization of SSA is different than the existing one in literature. A comparison
of our results with those obtained by other method is made.

\vskip 5mm
\noindent
\end{abstract}
\vskip 1cm

\par\noindent
{\bf 1. Introduction}
\par\vskip10pt
In scattering processes with a transversely polarized hadron in the initial state,
Single transverse-Spin Asymmetry(SSA) relative to the spin direction can be nonzero.
SSA has been observed in various experiments,
a review about the phenomenologies can be found in \cite{Review}.
Theoretically, SSA can be predicted with the concept of QCD factorization,
if large momentum transfers are large.
In the factorization of SSA
nonperturbative effects of the transversely polarized hadron
are factorized into matrix elements of the hadron. Therefore,
it will provide a new way to study the inner-structure of hadron by studying SSA. 
In this work we study the collinear factorization of SSA in Drell-Yan processes.
\par
From general principles SSA can be generated if the strong interaction
changes the helicities of hadrons in a scattering and the scattering amplitude
has an absorptive part. In the scattering involving a transversely polarized heavy quark, the helicity
of which is not conserved in QCD because of the heavy mass.  The related SSA can be calculated
with perturbative theory of QCD, e.g., in \cite{KPR,DGBB}.
For light hadrons in high energy processes, one can neglect
the mass of light quarks. The helicity of a massless quark is  conserved in QCD.
But this does not mean that the helicity of a light hadron is conserved in QCD,
because a light hadron is a  bound state of light quarks and gluons and the helicity 
of a light hadron is not only a sum of helicities  of light quarks. 

\par
The collinear factorization for describing SSA has been proposed in \cite{QiuSt, EFTE}.
With the collinear factorization SSA in various processes has been studied
in \cite{KaKo,EKT,KKnew,tw31,tw32,JQVY1,JQVY2,KVY}.
In such a factorization, the nonperturbative effects of the transversely polarized hadron
are factorized into twist-3 matrix elements,
or called ETQS matrix elements. Taking Drell-Yan processes as an example,
SSA is factorized as a convolution of three parts: The first part is the standard parton distribution
function of the unpolarized hadron defined with twist-2 operators. The second part consists of
matrix elements of the polarized hadron defined with twist-3 operators.
The third part consists of perturbative coefficient functions describing the hard scattering of partons.
If the factorization can be proven, the coefficient
functions are free from any soft divergence
like collinear- and I.R. divergence.
In this approach the effects of helicity-flip
are parameterized with twist-3 matrix elements, while the absorptive part of the scattering amplitude is generated
in the hard scattering of partons.
\par
\par
\begin{figure}[hbt]
\begin{center}
\includegraphics[width=5cm]{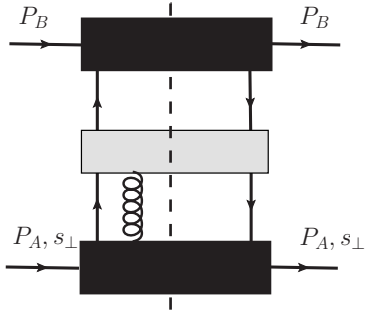}
\end{center}
\caption{The cut diagram for the differential cross section
of hadron scattering where the lower hadron is transversely polarized.
The broken line is the cut.
The black boxes represent parton density matrices of corresponding hadrons,
the gray box is the forward scattering amplitude of partons.}
\label{Fdg3}
\end{figure}
\par
A widely used method to derive the factorization of of SSA at the leading order of $\alpha_s$ 
is the diagram expansion at hadron level. All existing results are derived with the method except those 
in \cite{MS1,MS2,MS3,MS4}. This method has been also used for analyzing higher-twist effects, e.g., in \cite{HW1,HW2}. 
We take the Drell-Yan process $h_A + h_B \to \ell^+ \ell^- +X$
as an example to illustrate the method. In the process $h_A$ is transversely polarized
with the spin vector $s_\perp$. The spin-dependent part of the differential
cross section can be given by Fig.1. In Fig.1., the gray box represents
Feynman diagrams for various contributions of forward parton-scattering with the cut.  
The lower black box represents the density matrix of the polarized
hadron defined with quark- and gluon fields, and the upper black box
represents the quark density matrix of the unpolarized hadron. The three parts 
are connected with parton lines.  
A collinear expansion of the momenta carried by the parton lines 
is performed to pick up the leading power contributions. For the partons from $h_A$ the momentum is expanded around $P_A$,  while
the expansion for the partons from $h_B$ is around $P_B$.
After the expansion, one obtains an approximated form of density matrices parameterized with
various nonperturbative functions, i.e., parton distributions functions and twist-3 matrix elements,
and the perturbative coefficient functions of the factorization.
It is interesting to note that SSA in the factorization contains not only the so-called hard-pole contributions
in which all three patrons
from $h_A$ carry nonzero momentum fraction, but also the so-called soft-pole contributions
in which one of the three partons can have zero momentum fraction.
\par
It should be noted that QCD
factorizations, if they are proven, are general properties of QCD.
These factorizations hold not only at hadron level but also when one replaces
the hadron states with parton states. 
The perturbatively calculable parts in factorizations, i.e., the perturbative coefficient functions, do not depend
on hadrons and  are completely determined by the scattering of
partons.
To derive the factorization of SSA, one can replace hadrons with parton states and calculate
SSA perturbatively. The relevant twist-2 and twist-3 matrix elements can also be calculated 
with the parton states. 
In general the obtained results will contain soft divergences which usually appear
beyond leading order. By writing SSA as a convolution
of these matrix elements and perturbative coefficient functions, one can determine
the functions. In this work we will take this approach to derive at leading order all perturbative coefficient 
functions appearing in the collinear factorization of SSA in Drell-Yan processes. 
\par 
We notice that the approach taken here has been used to study factorizations only involving twist-2 operators. 
Applying the approach for SSA, i.e., factorizations with twist-3 operators, will provide 
an alternative way to derive the factorization or to calculate perturbative coefficient functions. 
This will also give an independent verification of results derived with other approaches.
It is not the intention here to prove the factorization or that the perturbative coefficient functions
are free from soft divergences at any order. This is beyond the scope of the present work. However, 
as we will see, we already have at leading order some perturbative coefficient functions obtained by subtraction 
of collinear divergences with twist-3 matrix elements, in contrast to the case only with twist-2 operators.  
  
\par
For the factorization only involve twist-2 operators, e.g., for the unpolarized differential cross-section
one can simply replace each hadron state with a single parton state
to derive the factorization.
But for SSA, because of the helicity conservation of QCD it is not possible to obtain nonzero SSA and the relevant twist-3 matrix elements by replacing the transversely polarized
hadron with a single quark state .
But, one can construct multi-parton states to replace the polarized hadron. With the multi-parton states
SSA and relevant twist-3 matrix elements are nonzero, because the helicity-flip effects can be generated through
correlations between these partons.
\par
In \cite{MS1,MS2} we have used
multi-parton states to study the factorization. We have found \cite{MS1} that with tree-level results of SSA and twist-3 matrix elements there are only the hard-pole contributions. Later, in \cite{MS2}
it has been realized that there is a special class of one-loop contributions to SSA which can
not be factorized as one-loop corrections to the hard-pole contributions at tree-level.
These one-loop
contributions can only be factorized with some special twist-3 matrix elements at one-loop.
These contributions are just the so-called soft-pole contributions.
Their perturbative functions, although extracted from parton scattering amplitudes at one-loop, are at the same
order as the hard-pole contributions derived from tree-level amplitudes.
\par
In this work we will use multi-parton states to derive all contributions in the factorization
formula for SSA in Drell-Yan processes. They are Hard-Pole(HP) contributions,
Soft-Quark-Pole(SQP) contributions and Soft-Gluon-Pole(SGP) contributions. There are two types
of SGP contributions. One is from the case as given in Fig.1, where the gluon 
from the polarized hadron has zero momentum fraction. Another is from the case where  three gluons
are from the polarized hadron  
and one of them carries zero momentum.
The twist-3 matrix elements for the three gluon case have been defined in \cite{Ji3G}.
\par
It should be mentioned that besides the collinear factorization, there is another 
factorization for SSA in limited regions of kinematics.
If the transverse momentum of the lepton pair is small, one can use the Transverse-Momentum-Dependent(TMD) factorization.The TMD factorization for unpolarized cases have been studied in
\cite{CS,CSS,JMY,JMYG,CAM}. For SSA
nonperturbative effects of the polarized hadron are factorized into
Sivers function\cite{Sivers}. The properties of Sivers function  and SSA with it
have been studied extensively
\cite{JC,SJ1,TMDJi,Mulders97,Boer03,Anselmino,Mulders,DeSanctis,Efremov,BQMa}.
In \cite{MS1,MS3,MS4}
we have also examined the TMD factorization of SSA  with parton states
and found an agreement with existing results.
\par
Our work is organized as the following: In Sect.2 we give our notations
for Drell-Yan processes and the definitions of relevant
twist-3 matrix elements.  In Sect.3 we introduce
our multi-parton states. With these states one can define corresponding
spin-density matrices in helicity space. The non-diagonal parts of the matrices are relevant
for calculating SSA and twist-3 matrix elements. In Sect.4 we study SSA in the scattering
of multi-parton state at tree-level. With tree-level results we can derive HP contributions.
In Sect.5 and Sect.6 we consider SSA at one-loop level. We find a special class of one-loop
contributions which give the SQP- and SGP contributions.
Sect.7 is our summary.

\par\vskip20pt
\noindent
{\bf 2. SSA in Drell-Yan Processes and Definitions of Twist-3 Matrix Elements}
\par\vskip10pt
We will use the  light-cone coordinate system, in which a
vector $a^\mu$ is expressed as $a^\mu = (a^+, a^-, \vec a_\perp) =
((a^0+a^3)/\sqrt{2}, (a^0-a^3)/\sqrt{2}, a^1, a^2)$ and $a_\perp^2
=(a^1)^2+(a^2)^2$. Other notations are:
\begin{equation}
  g_\perp^{\mu\nu} = g^{\mu\nu} - n^\mu l^\nu - n^\nu l^\mu,
  \ \ \ \ \ \
  \epsilon_\perp^{\mu\nu} =\epsilon^{\alpha\beta\mu\nu}l_\alpha n_\beta, \ \ \ \
  \epsilon^{\alpha\beta\mu\nu} = -\epsilon_{\alpha\beta\mu\nu},  \ \ \ \ \
  \epsilon^{0123}=1
\end{equation}
with the light-cone vectors $l$ and $n$ defined as $l^\mu=(1,0,0,0)$ and $n^\mu=(0,1,0,0)$, respectively.
We consider the Drell-Yan process:
\begin{equation}
  h_A ( P_A, s) + h_B(P_B) \to \gamma^* (q) +X \to  \ell^-  + \ell ^+  + X,
\end{equation}
where $h_A$ is a spin-1/2 hadron with the spin-vector $s$.
We take a light-cone coordinate system in which the momenta and the spin are :
\begin{equation}
P_{A}^\mu \approx (P_{A}^+, 0, 0,0),  \ \ \ \ P_{B}^\mu \approx  (0,P_{B}^-, 0,0),  
\ \ \ \ s^\mu =(0,0, \vec s_\perp).
\end{equation}
The mass of hadrons are neglected.  The spin of $h_B$
is averaged. The invariant mass of the observed lepton pair is $Q^2 =q^2$.
We are interested in the spin-dependent part of the differential cross section,
which can be written as:
\begin{equation}
\frac{ d\sigma }{  d^2 q_\perp d q^+ d q^- }(\vec s_\perp)
 -\frac{ d\sigma }{  d^2 q_\perp d q^+ d q^- }(-\vec s_\perp)
 = \frac{8\pi \alpha_{em}^2 }{3 S Q^2}
     \epsilon_\perp^{\alpha\beta}s_{\perp\alpha}{q_{\perp\beta}} {\mathcal W}_T,
\end{equation}
in which $S=2P_A^+ P_B^-$. We parameterize the momentum of the lepton pair as:
\begin{equation}
   q^\mu = (x P_A^+, y P_B^-, \vec q_\perp).
\end{equation}
The structure function ${\mathcal W}_T(x,y,q_\perp)$ is related to the spin dependent part of
the hadronic tensor
\begin{equation}
W^{\mu\nu}  = \sum_X \int \frac{d^4 x}{(2\pi)^4} e^{iq \cdot x} \langle h_A (P_A, s_\perp), h_B(P_B)  \vert
    \bar q(0) \gamma^\nu q(0) \vert X\rangle \langle X \vert \bar q(x) \gamma^\mu q(x) \vert
     h_B(P_B),h_A (P_A, s_\perp)  \rangle,
\label{Had-T}
\end{equation}
by:
\begin{equation}
 \left (-g_{\mu\nu} + \frac{q_\mu q_\nu}{q^2} \right ) W^{\mu\nu}
   = \epsilon_\perp^{\alpha\beta}s_{\perp\alpha}{q_{\perp\beta}} {\mathcal W}_T + \cdots,
\end{equation}
where $\cdots$ stand for spin-independent part.

\par
For large $Q^2$ the structure function ${\mathcal W}_T$ can be factorized in the form
of a convolution of perturbative functions with the standard parton distribution functions
of $h_B$ and twost-3 matrix elements of $h_A$. The definitions of standard parton distribution functions with twist-2 operators can be found in literature. Here we discuss the definitions of twist-3 matrix elements  
of the transversely polarized hadron. 
The quark-gluon twist-3 matrix elements have been introduced in \cite{QiuSt,EFTE} firstly.
We take a variant form and define them in the light-cone gauge $n\cdot G=0$ :
\begin{eqnarray}
 T_{q+} (x_1,x_2) & =& s_{\perp\mu} \int \frac{dy_1 dy_2}{4\pi}
   e^{ -iy_2 (x_2-x_1) P^+ -i y_1 x_1 P^+ }
 \langle P, \vec s_\perp \vert
           \bar\psi (y_1n )
\nonumber\\
       && \ \ \ \ \ \    \cdot \gamma^+ \left ( \tilde G^{+\mu}(y_2n) + i \gamma_5  G^{+\mu}(y_2n) \right )
         \psi(0) \vert P,\vec s_\perp \rangle ,
\nonumber\\
  T_{q-} (x_1,x_2) & =& s_{\perp\mu} \int \frac{dy_1 dy_2}{4\pi}
   e^{ -iy_2 (x_2-x_1) P^+ -i y_1 x_1 P^+ }
 \langle P, \vec s_\perp \vert
           \bar\psi (y_1n )
\nonumber\\
       && \ \ \ \ \ \    \cdot  \gamma^+ \left ( \tilde G^{+\mu}(y_2n) - i \gamma_5  G^{+\mu}(y_2n) \right ) \psi(0) \vert P,\vec s_\perp \rangle ,
\label{def-tw3q}
\end{eqnarray}
with  $\tilde G^{+\mu} =\epsilon_\perp^{\mu\nu} G^{+}_{\ \ \nu}$.
In other gauges gauge links along the direction $n$ should be added to make the definitions
gauge-invariant.
One can also use the projection $\gamma^+$ or $\gamma^+ \gamma_5$ to defined twist-3 matrix element
$T_{qF}(x_1,x_2)$ and $T_{q\Delta,F}(x_1,x_2)$,respectively,  as in \cite{QiuSt}. The relations
between these twist-3 matrix elements are:
\begin{equation}
T_{q+}(x_1,x_2) = T_{qF}(x_1,x_2) + T_{q\Delta,F}(x_1,x_2), \ \ \
T_{q-}(x_1,x_2) = T_{qF}(x_1,x_2) - T_{q\Delta,F}(x_1,x_2).
\end{equation}
One can show that the function $T_{qF}(x_1,x_2)$ is symmetric in $x_1$ and $x_2$ and
$T_{q\Delta,F}(x_1,x_2)$ is anti-symmetric in $x_1$ and $x_2$.

\par
The twist-3 matrix elements $T_{q\pm}(x_1,x_2)$ with $x_{1,2} > 0$ describe the correlation of those partons 
from $h_A$ ,
which enter a hard scattering, e.g., the gray part of Fig.1. 
In the hard scattering, the initial quark carries the momentum faction $x_2$,
the gluon carries the momentum fraction $x_1-x_2$ and the final quark carries the momentum fraction $x_1$.
If the gluon
momentum fraction is $x_1-x_2=0$,
the corresponding hard scattering introduces the SGP contribution to
SSA. If a quark carry zero momentum, i.e., $x_1=0$ or $x_2=0$,
the corresponding hard scattering introduces the SQP contribution to
SSA. It is clear that the SGP contributions are related to $T_{q+}(x,x)$ with
$T_{q+}(x,x) =T_{q-}(x,x)=T_{qF}(x,x)$,
while the SQP contributions are related to $T_{q\pm}(0,x)$ or $T_{q\pm}(x,0)$.
There are contributions with nonzero $x_{1,2}$ and $x_1\neq x_2$. These contributions are
HP contributions. For the case $x_1 <0$ or $x_2<0$ the corresponding quark fields in the definition 
represent antiquarks.
\par
Instead of two quarks combined one gluon entering the hard scattering, there can be three
gluons entering the hard scattering\cite{Ji3G}. The corresponding contributions can be factorized
with matrix elements defined with twist-3 gluonic operators. In this case, as we will shown,
there is a leading contribution of $\alpha_s$ in the factorization of SSA.
The contribution is a SGP contribution in which one of the three gluons carries
zero momentum fraction. In general there are two types of twist-3 gluonic operators, distinguished
by the color structure. One can define them in the gauge $n\cdot G=0$:
\begin{eqnarray}
O^{\alpha\beta\gamma} (x_1,x_2) &=& \frac{ g_s}{P^+}  d^{bca} \int\frac{dy_1 d y_2}{4\pi}
  e^{-i y_1 x_1 P^+  -i y_2 (x_2-x_1) P^+ }
\nonumber\\
   &&  \ \ \ \ \ \langle P,s_\perp \vert G^{b,+\beta}(y_1 n ) G^{c,+\gamma} (y_2n) G^{a,+\alpha}(0)
\vert P,s_\perp\rangle,
\nonumber\\
 N^{\alpha\beta\gamma}(x_1,x_2) &=& i\frac{g_s}{P^+} f^{bca} \int\frac{dy_1 dy_2 }{4\pi}
  e^{-i y_1 x_1 P^+ - i y_2 (x_2-x_1) P^+ }
\nonumber\\
  &&  \ \ \ \ \  \langle P,s_\perp \vert  G^{b,+\beta}(y_1 n ) G^{c,+\gamma} (y_2n) G^{a,+\alpha}(0)
\vert P,s_\perp\rangle,
\end{eqnarray}
There are two scalar functions can be defined for each type of color structure
in the case of $x_1=x_2=x$ for SGP contributions:
\begin{eqnarray}
  O^{\alpha\beta\gamma}(x,x)
&& =  g^{\alpha\beta}_\perp \tilde s^\gamma  x G_{d1} (x)
   + \left [ g_\perp^{\alpha\gamma}\tilde s^\beta + g_\perp^{\beta\gamma}\tilde s^\alpha
      \right ] x  G_{d2} (x),
\nonumber\\
  N^{\alpha\beta\gamma} (x,x) &&=
  g^{\alpha\beta}_\perp \tilde s^\gamma  x G_{f1} (x)
   + \left [ g_\perp^{\alpha\gamma}\tilde s^\beta + g_\perp^{\beta\gamma}\tilde s^\alpha
      \right ] x  G_{f2} (x).
\label{GF-def}
\end{eqnarray}
\par
In general the function ${\mathcal W}_T$ in the collinear factorization can be divided into four
parts:
\begin{equation}
  {\mathcal W}_T = {\mathcal W}_T\biggr\vert_{HP} + {\mathcal W}_T\biggr\vert_{SQP}
   + {\mathcal W}_T\biggr\vert_{SGPF}  + {\mathcal W}_T\biggr\vert_{SGPG} .
\label{SGP-GG}
\end{equation}
For SGP contributions one can have two different contributions factorized
with the quark-gluon- or purely gluonic twist-3 matrix elements, denoted  by
the subscriber $SGPF$ and $SGPG$ respectively.
Each of the four parts can be expressed as convolutions of parton distributions, twist-3 matrix elements
discussed in the above, and perturbative coefficient functions. Details of
the convolutions will be given in the following sections.
\par
The goal of our work is to derive all perturbative functions at leading order of $\alpha_s$.
For HP contributions at leading order we only need to calculate with parton states 
parton scattering amplitudes and twist-3 matrix elements at tree-level. 
At one-loop level, there are in general collinear
divergences in ${\mathcal W}_T$. As observed in \cite{MS2},   
at one-loop there is a class of contributions, whose collinear divergences can not be subtracted 
by using one-loop results of twist-3 matrix elements in the factorized HP contributions derived 
with tree-level results. These contributions hence can not be taken as one-loop corrections 
to the perturbative coefficient functions in HP contributions.  
In fact, the collinear divergences can be subtracted by the so-called soft-pole twist-3 matrix elements in which 
one parton carries zero momentum fraction. This is the origin of 
the soft-pole contributions. 
The soft-pole matrix elements are zero at tree-level, but nonzero 
at one-loop.  
This results in that the perturbative coefficient functions of the soft-pole contributions 
are at the same order of those of the HP contributions derived from tree-level.

\par\vskip20pt
\noindent
{\bf 3. Spin-Density Matrices and  Multi-parton States}
\par\vskip10pt
We consider a system $\vert N[\lambda]\rangle$ with total spin $1/2$.
The system moves in the $z$-direction with the helicity $\lambda=\pm$ and can be a superposition of various multi-parton states. 
We consider a forward scattering of the system through some operator ${\mathcal O}$ which do not change helicity of quarks. The operator can be
those used to define twist-3 matrix elements, or the hadronic tensor. In the later,
the forward scattering is with some additional particles which are unpolarized.
The
transition amplitude is given as:
\begin{equation}
 {\mathcal M}_{\lambda_2 \lambda_1} = \langle  N[\lambda_2 ] \vert {\mathcal O} \vert N [\lambda_1 ]\rangle.
\label{FW}
\end{equation}
We use $\lambda_{1,2}=\pm  $ to denote the helicity of the
initial- and final state, respectively. The transition amplitude in the helicity space is
$2\times 2$ matrix and can be expanded as:
\begin{equation}
{\mathcal M}_{\lambda_2 \lambda_1} = \left [ a + \vec b \cdot \vec \sigma \ \right ]_{\lambda_2 \lambda_1 },
\label{nondm}
\end{equation}
or it can also be described with a spin vector $s^\mu=(s^0,\vec s)$: 
\begin{equation}
{\mathcal M} (s)  = \left [ a + \vec b \cdot \vec s \  \right ],\ \ \ \ \ s^2 =-1.
\end{equation}
From the above the transverse-spin dependent part is determined by $\vec b_\perp$, i.e., the non-diagonal
part in the helicity space.  
SSA appears if the non-diagonal part of the hadronic tensor in Eq.(\ref{Had-T}) in the helicity space is non zero.
\par
Because of helicity-conservation of QCD
the non-diagonal part of ${\mathcal M}$ in Eq.(\ref{nondm}) is zero,   if $\vert N [\lambda ] \rangle$ 
is a single quark state.
Instead of a single quark one can consider the following multi-parton state:
\begin{equation}
 \vert N [\lambda ] \rangle  =  \vert q [\lambda ] \rangle + c_1
                   \vert q g [\lambda ] \rangle + c_2 \vert q q \bar q [\lambda ] \rangle +c_3\vert q gg [\lambda ] \rangle                  + \cdots ,
\label{past}
\end{equation}
where all partons move in the $z$-direction, the sum of helicities of partons is $1/2$ or $-1/2$. We will give
later the details about the momenta and color structure of these partons. The $\cdots$ stand for possible states
with more than 3 partons. We do not need to consider
the states with more than 3 partons. The reason is the following: The leading power contributions to SSA
come from those parton scattering processes, in which only three patrons from the polarized hadrons are
involved. We call these involved partons active partons.  Certainly, there can be more than three partons
as active partons. The resulted contributions
are power-suppressed and may be factorized with operators whose twist is larger than 3.
In the leading power contributions, the three active partons can be the combinations
of quarks and gluons. They are $qqg$, $q\bar qg$ and $ggg$. Charge-conjugated
combinations should also be included.  In the case described by 
Fig.1. the active partons are $qqg$. It is clearly that with first three states in Eq.(\ref{past}) one can have all combinations by taking some partons as spectators.
\par
If we calculate the non-diagonal part of ${\mathcal M}$ in Eq.(\ref{nondm}) with
the multi-parton state in Eq.(\ref{past}),
one will find with the helicity conservation of quarks nonzero contributions only from the interference
between different states in the right hand side of Eq.(\ref{past}).
If we replace $h_A$ with the above state and $h_B$ with a single unpolarized parton in Eq.(\ref{Had-T}),
we will also get nonzero result for the spin-dependent part of $W^{\mu\nu}$ or for ${\mathcal W}_T$.
Similarly, the defined twist-3 matrix elements are also nonzero with the state $ \vert N [\lambda ] \rangle$.
These nonzero results allow us to study the factorization of SSA.
\par
In \cite{MS3,MS4} factorizations of SSA in Drell-Yan processes have been studied
with the first two terms in Eq.(\ref{past}) in the kinematical region $q^2_\perp/Q^2\ll 1$. In this case, all partons are active. Non of them can be a spectator parton. But for interferences between other states,
some partons can be spectators, because we only need to consider
those interferences with three active partons.
The existence of possible spectators only affects overall factors of interested quantities like
${\mathcal W}_T$ and twist-3 matrix elements, it has no effect on the derivation of perturbative coefficient functions.
Hence, for our purpose we only need to consider
those matrix elements $\langle a b \vert {\mathcal O} \vert c\rangle$
or $\langle c \vert {\mathcal O} \vert  a b \rangle$, where $a,b$ and $c$ are quarks or gluons.
These matrix elements can be obtained from the interferences between different states in Eq.(\ref{past})
by taking out some partons as spectators.
We will illustrate this in the following.
\par
For the interference between the $q$- and the $qg$-component, non of partons can be a spectator.
We define the state $\vert q [\lambda ] \rangle$ and the state $ \vert q g [\lambda ] \rangle$
as:
\begin{eqnarray}
&& \vert  q (p,\lambda_q) \rangle = b^\dagger_{i_c} (p,\lambda_q) \vert 0 \rangle,
\ \ \ \ \
\vert q (p_1,\lambda_q)g(p_2,\lambda_g) \rangle =
T^a_{j_c i_c} b^\dagger_{j_c} (p_1,\lambda_q) a^\dagger_a (p_2,\lambda_g) \vert 0 \rangle,
\nonumber\\
  && p^\mu =(p^+, 0,0,0), \ \ \  p_1^\mu =x_0 p^\mu, \ \ \ \  p_2 =(1-x_0) p^\mu =\bar x_0 p^\mu,
\end{eqnarray}
where $b^\dagger_i$ is the quark creation operator with $i$ as the color index,
$a^\dagger_a$ is the gluon creation operator with $a$ as the color index.
$\lambda_q(\lambda_g)$ is the helicity of the quark(gluon). To simplify the notations we will
write $p^+ =P_A^+$ and $\bar p^- = P_B^-$.
It is straightforward to obtain the non-diagonal part as
\begin{eqnarray}
{\mathcal M}^{(qg)}_{+ - } &=& {\mathcal C}^{qg}\biggr [ \langle q(p,+)  \vert {\mathcal O} \vert q(p_1,+ ) g(k, - )  \rangle
                        + \langle q(p_1,- ) g(k, +  )  \vert {\mathcal O} \vert q(p,-) \rangle \biggr ],
\nonumber\\
{\mathcal M}^{(qg)}_{- + } &=& {\mathcal C}^{qg} \biggr [ \langle q(p,-)  \vert {\mathcal O} \vert q(p_1,- ) g(k, + )  \rangle
                        + \langle q(p_1,+ )g(k,-)  \vert {\mathcal O} \vert q(p,+) \rangle \biggr ].
\label{ndmqg}
\end{eqnarray}
In the above
we use the index $qg$ to denote this type of interference contribution.
We also introduce a coefficient ${\mathcal C}^{qg}$ in the non-diagonal part
of the spin-density matrix. Contributions of this type to  twist-3 matrix elements and
${\mathcal W}_T$ will be proportional to ${\mathcal C}^{qg}$, and will be called $qg$-contributions.
The derived perturbative function will not depend on  ${\mathcal C}^{qg}$.

\par
For the contribution from the $qq\bar q$-state we note that the interference with the single quark state
is zero if the total helicity is changed. The interference with the $qgg$-state does not need
to be considered because at least four partons must be active and one quark is a spectator.
Hence, we only need to consider the interference with the $qg$-state, in which one quark is a spectator and three
partons are active.
In this case, the forward scattering is participated by a gluon and a $q\bar q$-pair.
In order to have $\Delta \lambda =\pm 1$, the total helicity $\lambda$ of the $q\bar q$-pair must be zero.
There can be two states with $\lambda=0$ for the $q \bar q$-pair. We denote the two states as :
\begin{eqnarray}
\vert \left ( q \bar q \right ) _\pm \rangle  = T^a_{j_c i_c} \left [  b^\dagger_{j_c} (p_1,+) d^\dagger_{i_c} (p_2,-)
 \pm  b^\dagger_{j_c} (p_1,- ) d^\dagger_{i_c} (p_2,+ ) \right  ] \vert 0 \rangle,
\end{eqnarray}
and the single gluon state as $\vert g(\lambda_g)\rangle$. The gluon carries the same color index $a$ and the momentum
$p$. $\lambda_g$ is the helicity.
$d^\dagger$ is the create operator for the antiquark. With these states one can construct the non-diagonal
part of the spin-density matrix ${\mathcal M}^{(q\bar q)}$ as:
\begin{eqnarray}
{\mathcal M}^{(q\bar q)}_{+-} &=&  {\mathcal C}^{q\bar q}_{+}
\biggr [ \langle \left ( q \bar q \right )_+ \vert {\mathcal O} \vert g(- )  \rangle
                        + \langle  g(+ ) \vert {\mathcal O} \vert \left ( q \bar q \right )_+ \rangle \biggr ]
                        +{\mathcal C}^{q\bar q} _{-}
\biggr [ \langle \left ( q \bar q \right )_- \vert {\mathcal O} \vert g( - )  \rangle
                        - \langle  g(+ ) \vert {\mathcal O} \vert \left ( q \bar q \right )_- \rangle \biggr ],
\nonumber\\
{\mathcal M}^{(q\bar q)}_{-+} &=&  {\mathcal C}^{q\bar q} _{+}
\biggr [ \langle \left ( q \bar q \right )_+ \vert {\mathcal O} \vert g(+ )  \rangle
                        + \langle  g(- ) \vert {\mathcal O} \vert \left ( q \bar q \right )_+ \rangle \biggr ]
                        - {\mathcal C}^{q\bar q}_{-}
\biggr [  \langle \left ( q \bar q \right )_- \vert {\mathcal O} \vert g( + )  \rangle
                        - \langle  g(- ) \vert {\mathcal O} \vert \left ( q \bar q \right )_- \rangle \biggr ].
\label{ndmd}
\end{eqnarray}
In the above we have introduced two coefficients ${\mathcal C}^{q\bar q}_\pm$
to distinguish the contributions from the two $q\bar q$
states. We note that there is a sign difference for the terms with ${\mathcal C}^{q\bar q}_-$ between the first and the second equation.  This difference can be easily
found by requiring that the state $\vert \left ( q\bar q \right )_- \rangle$ becomes a spin-1/2 system by adding
a quark. Again, we will call all contributions from this matrix as $q\bar q$-contributions. They
are linear in the two coefficients ${\mathcal C}^{q\bar q}_\pm$. The derived perturbative
functions will not depend on ${\mathcal C}^{q\bar q}_\pm$ .

\par
For the contribution from  the $qgg$-state in Eq.(\ref{past}), only the interference
with the $qg$-state and with the $qgg$-state need to be considered here, where one quark
can be taken as a spectator. In this case, one has the forward scattering as $ gg \to g$
or $g\to gg$. The color of the two gluon state must
be the same as the color of the one gluon state. The total helicity $\lambda$ of the two gluons must be zero.
There are two states with $\lambda=0$ for a given color structure. We denote
\begin{eqnarray}
 \vert (gg)_{\pm} \rangle = i f^{abc}  \left [  a_b^\dagger (p_1,+) a_c^\dagger (p_2,-)
 \pm  a_b^\dagger (p_1,- ) a_c^\dagger(p_2,+ ) \right  ] \vert 0 \rangle.
\label{ggF}
\end{eqnarray}
With these states one can construct the non-diagonal
element of the spin-density matrix ${\mathcal M}^{(ggF)}$ as:
\begin{eqnarray}
{\mathcal M}^{(ggF)}_{+-} &=&  {\mathcal F}^{gg}_{+}
\biggr [ \langle \left ( gg \right )_{+} \vert {\mathcal O} \vert g(- )  \rangle
                        + \langle  g(+ ) \vert {\mathcal O} \vert \left (gg \right )_{+ }\rangle \biggr ]
                        -{\mathcal F}^{gg} _{-}
\biggr [ \langle \left ( gg \right )_{-} \vert {\mathcal O} \vert g( - )  \rangle
                        - \langle  g(+ ) \vert {\mathcal O} \vert \left ( gg \right )_{-} \rangle \biggr ],
\nonumber\\
{\mathcal M}^{(ggF)}_{-+} &=&  {\mathcal F}^{gg} _{+}
\biggr [ \langle \left ( gg \right )_+ \vert {\mathcal O} \vert g(+ )  \rangle
                        + \langle  g(- ) \vert {\mathcal O} \vert \left ( gg \right )_+ \rangle \biggr ]
                        + {\mathcal F}^{gg }_{-}
\biggr [  \langle \left ( gg \right )_- \vert {\mathcal O} \vert g( + )  \rangle
                        - \langle  g(- ) \vert {\mathcal O} \vert \left ( gg  \right )_- \rangle \biggr ].
\label{ndmggF}
\end{eqnarray}
In the above we introduce two coefficients ${\mathcal F}^{gg}_\pm$ to distinguish the contributions
from the two states in Eq.(\ref{ggF}).
Another spin-density matrix ${\mathcal M}^{(ggD)}$ can be constructed in this case by replacing $if^{abc}$
in Eq.(\ref{ggF}) with $d^{abc}$, and ${\mathcal F}^{gg}_\pm$
with ${\mathcal D}^{gg}_\pm$ in Eq.(\ref{ndmggF}). We will call all contributions from
these two spin-density matrices as $gg$-contributions. They are linear in the four coefficients
${\mathcal F}^{gg}_\pm$ and ${\mathcal D}^{gg}_\pm$.
\par
With the constructed spin-density matrices in the above one can calculate twist-3 matrix elements
and the structure functions ${\mathcal W}_T$ by taking correspond operator ${\mathcal O}$.
It is straightforward
to obtain the twist-3 matrix elements $T_{q\pm}$ at tree-level.
The results from the $qg$-contributions are
\begin{eqnarray}
T_{q+} (x_1,x_2) &=& {\mathcal C}^{qg} \pi g_s \sqrt{2 x_0} (N_c^2-1)(x_2-x_1)
                      \delta (1-x_1) \delta (x_2-x_0 ) +{\mathcal O}(g_s^3),
\nonumber\\
T_{q-} (x_1,x_2) &=&  - {\mathcal C}^{qg} \pi g_s \sqrt{2 x_0 } (N_c^2-1)(x_2-x_1)
  \delta (1-x_2) \delta (x_1-x_0) +{\mathcal O}(g_s^3).
\label{tree-Tq}
\end{eqnarray}
The results from the $q\bar q$-contributions are
\begin{eqnarray}
T_{q+ } (x_1,x_2) &=&  \pi g_s (N_c^2-1) \sqrt{2x_0\bar x_0 } \biggr [ \left ({\mathcal C}^{q\bar q}_+ -{\mathcal C}^{q\bar q}_- \right )  \delta (x_1 +\bar x_0) \delta (x_2-x_0)
\nonumber\\
   && \ \ \ \ \ \ \ \ \ \  + \left ({\mathcal C}^{q\bar q}_+ + {\mathcal C}^{q\bar q}_- \right )
\delta (x_2 +\bar x_0) \delta (x_1-x_0) \biggr ] +{\mathcal O}(g_s^3),
\nonumber\\
T_{q-}  (x_1,x_2) &=&\pi g_s (N_c^2-1) \sqrt{2x_0\bar x_0 } \biggr [ \left ({\mathcal C}^{q\bar q}_+ +{\mathcal C}^{q\bar q}_- \right )  \delta (x_1 +\bar x_0) \delta (x_2-x_0)
\nonumber\\
   && \ \ \ \ \ \ \ \ \ \  + \left ({\mathcal C}^{q\bar q}_+ - {\mathcal C}^{q\bar q}_- \right )
\delta (x_2 +\bar x_0) \delta (x_1-x_0) \biggr ] +{\mathcal O}(g_s^3).
\end{eqnarray}
With ${\mathcal W}_T$ calculated in the next section at leading order, one can find
the factorized form of ${\mathcal W}_T$ in terms of $T_{q\pm}$ with the above results.
This is for HP contributions.
In Sect.5 and 6 we will also give the results of $T_{q\pm} (x,x)$, $T_{q\pm} (0,x)$
and these gluonic twist-3 matrix elements. These results are at order of $g_s^3$
and will be used to factorize the soft-pole contributions.

\par\vskip20pt
\noindent
{\bf 4. Hard-Pole Contributions}
\par\vskip10pt
As discussed in the last section, we replace the polarized hadron $h_A$ with the multi-parton state
in Eq.(\ref{past}) to calculate the non-diagonal part of the constructed spin-density
matrices for ${\mathcal W}_T$. We replace the unpolarized hadron $h_B$ with single-parton states.
In this section we will work at tree-level. 

\par
\begin{figure}[hbt]
\begin{center}
\includegraphics[width=9cm]{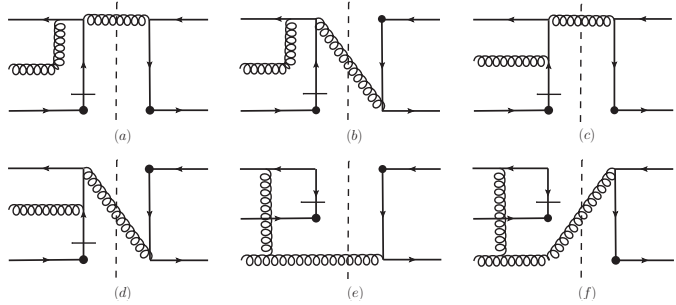}
\end{center}
\caption{The diagrams for the amplitude
$\bar q + (q + G) \to \gamma^*  + X \to \bar q + q $ at tree-level. The black dots denote
the insertion of electromagnetic current operator. Broken lines represent the cut. For the propagators
with a short bar only the absorptive part of the propagator is taken into account. }
\label{HP-P1}
\end{figure}

\par
We first consider the $qg$-contributions. If we replace $h_B$ with an antiquark $\bar q$ with the
momentum $\bar p^\mu =(0,\bar p^-,0,0)$,  the leading order contribution to ${\mathcal W}_T$ comes
from diagrams in Fig.2. The complex conjugated diagrams should
be included in order to obtain the non-diagonal part of the spin-density matrix ${\mathcal M}^{(qg)}$ given in
Eq.(\ref{ndmqg}). In the diagrams of Fig.2. the broken line divides each diagrams into a left- and right part.
Each part represents a scattering amplitude.  The short bar cutting a quark propagator is in fact
a physical cut of the amplitude represented by the left part. It means that only the absorptive part
of the cutting propagator is taken into account:
\begin{equation}
{\rm Abs} \left [ \frac{i \gamma\cdot k_q}{k_q^2 + i\varepsilon} \right ] = \pi \delta (k_q^2) \gamma\cdot k_q,
\end{equation}
It is straightforward to calculate these diagrams and we obtain:
\begin{eqnarray}
{\mathcal W}_T\biggr\vert_{Fig.\ref{HP-P1}} &=&  -{\mathcal C}^{qg} e_q^2 \frac{g_s \alpha_s}{4\pi}  \frac{\sqrt{2x_0}}{q^2_\perp}
     \frac{N_c^2-1}{N_c^2 } \delta(\bar x -y\bar x_0  ) \delta (s(1-x)(1-y)-q^2_\perp) (N_c^2 +y-1)
\nonumber\\
        && \cdot \frac{1}{1-y}  \left [ y^2 + x_0 -\vert \lambda_q \vert (y^2-x_0) \right ],
\label{qWT}
\end{eqnarray}
with $s =2 p^+ \bar p^-$. $e_q$ is the electric charge of the quark $q$ in unit $e$.
The $\delta$-function of $1-x-y\bar x_0 $ is from the cutting quark propagator.
The terms with $\vert \lambda_q \vert =1$ are quark-spin dependent, because the external quark lines
are extracted with $\lambda_q \gamma_5 \gamma\cdot p$.

\par
With the $qg$-contributions of $T_{q\pm}$ one can write the above ${\mathcal W}_T$ into a factorized form.
The terms with $\vert \lambda_q \vert =1$ should be factorized with $T_{q+}-T_{q-}$ or $T_{q\Delta, F}$, because
$\gamma^+\gamma_5$ is used to define them. The other terms should be factorized with $(T_{q+}+T_{q-})$ or $T_{qF}$.
With $T_{q\pm}$ in Eq.(\ref{tree-Tq}) we have the factorized form:
\begin{eqnarray}
{\mathcal W}_T\biggr\vert_{Fig.\ref{HP-P1}} & = &  \frac{e_q^2 \alpha_s}{\pi^2 N_c q^2_\perp }
\int_x^1 \frac{dy_1}{y_1} \int_y^{1} \frac{d y_2}{y_2} f_{\bar q} (y_2)\delta (\hat s (1-\xi_1)(1-\xi_2) -q^2_\perp)
\cdot \biggr  [ {\mathcal H}_{q+} (\xi_1,\xi_2) T_{q+} (y_1, x_B )
\nonumber\\
   && \  \ \ \ \ \ \  + {\mathcal H}_{q-} (\xi_1,\xi_2)   T_{q-} (y_1, x_B ) \biggr ],
\label{Fac-Fig2}
\end{eqnarray}
with
\begin{eqnarray}
 {\mathcal H}_{q+} (\xi_1,\xi_2) &=&  \frac{ N_c^2 +\xi_2 -1 }{2N_c\xi_2 (1-\xi_1)(1-\xi_2)}
 \left ( \xi_2+ \xi_1  -1 \right ),
\ \ \
 {\mathcal H}_{q-} (\xi_1,\xi_2) = \frac{ N_c^2 +\xi_2 -1 }{2 N_c (1-\xi_1)(1-\xi_2)} \xi_2^2,
\nonumber\\
 \xi_1 &=& \frac{x}{y_1},  \ \ \ \ \xi_2 =\frac{y}{y_2}, \ \ \ x_B = \frac{q^2}{2 q\cdot p} ,
  \ \ \ \ \hat s = y_1 y_2 s.
\end{eqnarray}
The function $f_{\bar q} (y_2)$ is the antiquark distribution function of $h_B$. For $h_B =\bar q$, we have $f_{\bar q} (y_2)
=\delta (1-y_2)+{\mathcal O}(\alpha_s)$. It is noted that the derived perturbative coefficient functions 
do not depend on ${\mathcal C}^{qg}$. 
One can also replace $h_B$ with a quark. In this case the results can be obtained
by reversing the directions of quark lines in Fig.2. They can be obtained from
the above results through charge-conjugation. We will give them at the end of this section
by combining all parton flavors.

\par
\begin{figure}[hbt]
\begin{center}
\includegraphics[width=9cm]{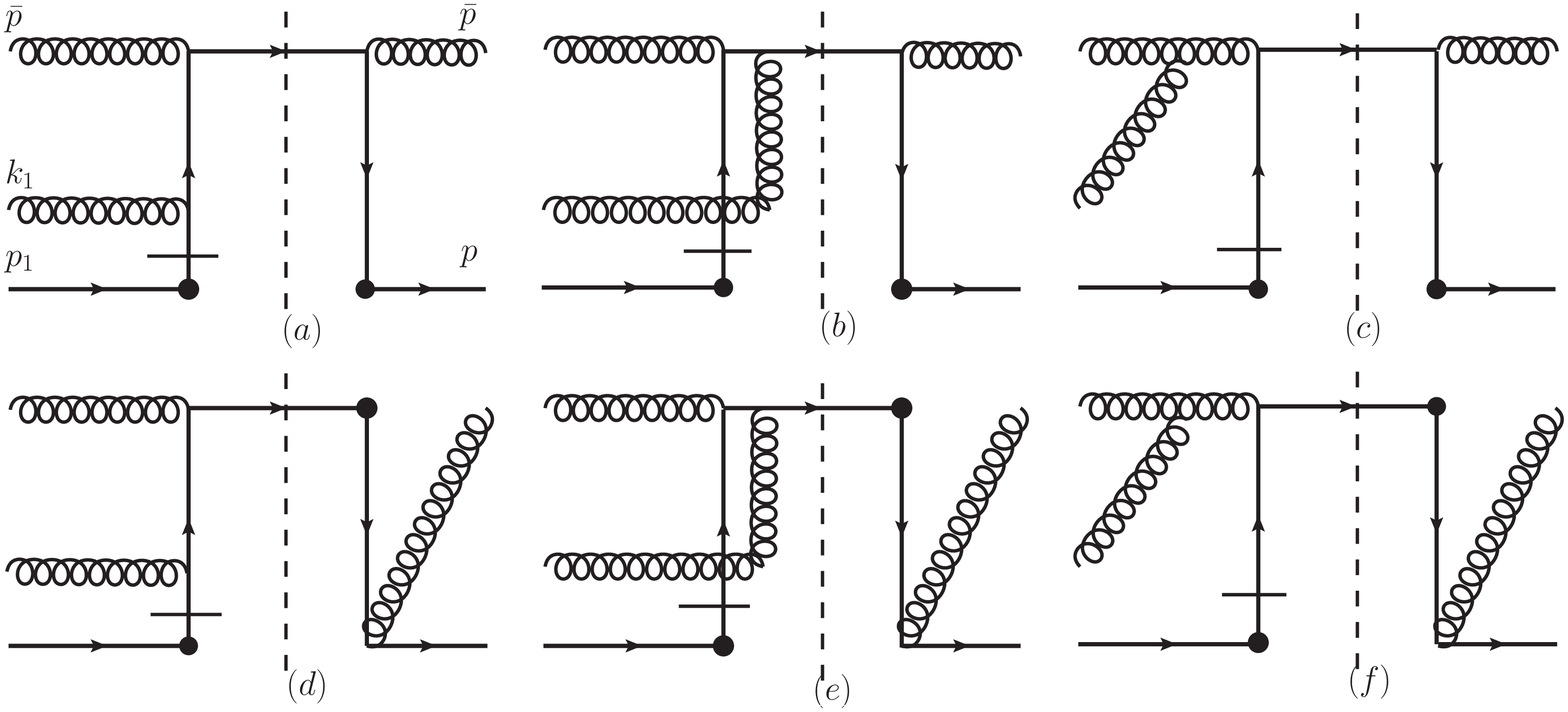}
\end{center}
\caption{The diagrams for the amplitude $G + \left [ q +G \right ] \to \gamma^* + X \to   G + q$ at tree-level}
\label{HP-P2}
\end{figure}
\par
If we replace the unpolarized hadron $h_B$ with a gluon carrying the momentum $\bar p$, the leading contributions to ${\mathcal W}_T$
comes from Fig.3. The calculation of these diagrams is similar to the calculation of Fig.2. We have
the sum of Fig.3:
\begin{eqnarray}
{\mathcal W}_T\biggr\vert_{Fig.\ref{HP-P2}} &=& {\mathcal C}^{qg} e_q^2 \frac{g_s \alpha_s}{4\pi N_c} \sqrt{2 x_0}
 \delta (s(1-x)(1-y)-q^2_\perp) \delta (\bar x -y\bar x_0)
\frac{1 + (y-1) N_c^2}{  q^2_\perp}
\nonumber\\
 &&  \cdot \biggr [ x_0 (1-y)^2 + y^2 + \vert \lambda_q \vert (x_0 (1-y)^2 -y^2 ) \biggr ].
\label{gWT}
\end{eqnarray}
This result can be factorized in the following form:
\begin{eqnarray}
{\mathcal W}_T\biggr\vert_{Fig.\ref{HP-P2}} &=& \frac{e_q^2\alpha_s }{\pi^2 N_c q^2_\perp }
 \int_x^1 \frac{dy_1}{y_1} \int_y^1\frac{dy_2}{y_2}
 f_g (y_2) \delta(\hat s(1-\xi_1)(1-\xi_2)-q^2_\perp)
\nonumber\\
 && \cdot \biggr [  {\mathcal H}_{g+}(\xi_1,\xi_2)  T_{q+} (y_1, x_B )
     +  {\mathcal H}_{g-}(\xi_1,\xi_2)  T_{q-} (y_1, x_B )  \biggr ] ,
\nonumber\\
  {\mathcal H}_{g+}(\xi_1,\xi_2) &= &\frac{1+ (\xi_2 -1)N_c^2}{ 2 (N_c^2-1) (1-\xi_1)\xi_2}
   (1-\xi_2)^2 (1-\xi_1-\xi_2),
\nonumber\\
   {\mathcal H}_{g-}(\xi_1,\xi_2) &=& -  \frac{1+ (\xi_2 -1)N_c^2}{ 2(N_c^2-1) (1-\xi_1)} \xi_2^2.
\label{qHP}
\end{eqnarray}
where $f_g (y_2)$ is the gluon distribution function. For $h_B=g$ we have $f_g (y)=\delta (1-y) + {\mathcal O}(\alpha_s)$.

\par
\begin{figure}[hbt]
\begin{center}
\includegraphics[width=13cm]{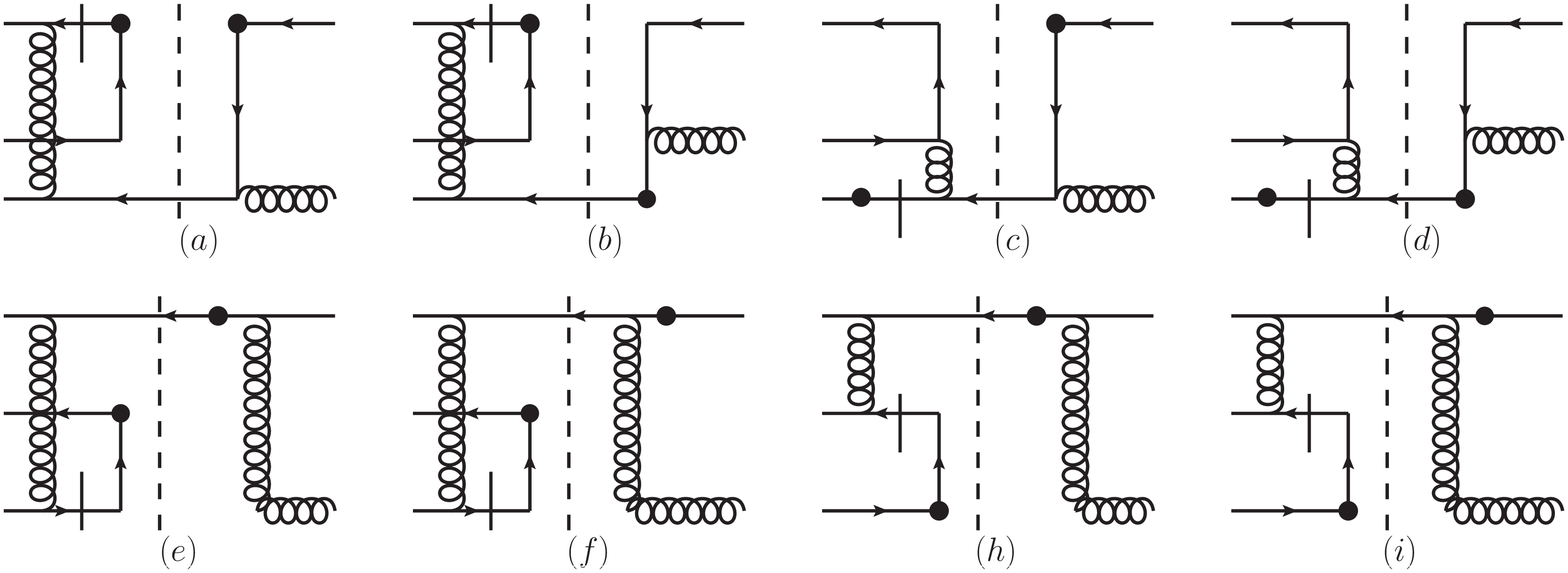}
\end{center}
\caption{The diagrams for the amplitude
$\bar q + \left [ q + \bar q \right ]  \to \gamma^*  + X \to \bar q + G $ at tree-level.  }
\label{HP-P3}
\end{figure}
\par
\par
By replacing $h_B$ with an antiquark $\bar q(\bar p)$, the $q\bar q$-contributions for ${\mathcal W}_T$
are also at leading order. They are given by the diagrams in Fig.4. In the first four diagrams
the anti-quark $\bar q(\bar p)$ in the initial single parton state must have the same flavor as the quark in the multi-parton state, while in the last four diagrams
$\bar q(\bar p)$ can have different flavor.
The results are:
\begin{eqnarray}
{\mathcal W}_{T} \biggr\vert_{\ref{HP-P3}a + \ref{HP-P3}b+ \ref{HP-P3}c + \ref{HP-P3}d } &=& e_q^2 \frac{g_s\alpha_s}{2\pi q_\perp^2}
 \sqrt{2x_0\bar x_0} \frac{N_c^2-1}{N_c^2} \left ( {\mathcal C}^{q\bar q}_- + {\mathcal C}^{q\bar q}_+ (1-2x_0)\right )
  \delta (s (1-x)(1-y)-q_\perp^2)
\nonumber\\
   && \ \ \ \cdot y \left [ \frac{\delta (\bar x -y\bar x_0) }{1-y} -(1-y)^2 \delta (\bar x-y x_0)
    \right ],
\nonumber\\
 {\mathcal W}_{T} \biggr\vert_{\ref{HP-P3}e + \ref{HP-P3}f+ \ref{HP-P3}h + \ref{HP-P3}i } &=&
- e^2_q \frac{g_s\alpha_s}{2\pi q^2_\perp}
 \sqrt{2x_0\bar x_0} \frac{N_c^2-1}{N_c} \left ( {\mathcal C}^{q\bar q}_- + {\mathcal C}^{q\bar q}_+ (1-2x_0)\right )  \delta (s (1-x)(1-y)-q_\perp^2)
\nonumber\\
&& \ \ \cdot  \biggr [ \delta (\bar x -y\bar x_0)  +\delta (x -y\bar x_0)
    \biggr ] \left ( y^2-2y+2 \right ) .
\end{eqnarray}
With the tree-level results of the $q\bar q$-contributions for the twist-3 matrix elements we can derive the following
factorized form:
\begin{eqnarray}
{\mathcal W}_{T} \biggr\vert_{\ref{HP-P3}a + \ref{HP-P3}b+ \ref{HP-P3}c + \ref{HP-P3}d }
&=& \frac{ e_q^2\alpha_s}{\pi^2 N_c q^2_\perp}
\int_x^1 \frac{dy_1}{y_1} \int_y^1 \frac{dy_2}{y_2} f_{\bar q} (y_2) \delta (\hat s (1-\xi_1)(1-\xi_2) -q^2_\perp)
\nonumber\\
  && \cdot \biggr\{  {\mathcal H}_{q\bar q +} (\xi_1,\xi_2) T_{q+}(-\hat y_1, x_B)
     + {\mathcal H}_{q \bar q -}(\xi_1,\xi_2) T_{q-}(-\hat y_1, x_B )
\nonumber\\
  && + \biggr [ {\mathcal H}_{\bar q q +} (\xi_1,\xi_2) T_{q+}(-x_B, \hat y_1)
   + {\mathcal H}_{\bar q q -} (\xi_1,\xi_2) T_{q-}(-x_B, \hat y_1 ) \biggr ] \biggr \} ,
\nonumber\\
{\mathcal W}_{T} \biggr\vert_{\ref{HP-P3}e + \ref{HP-P3}f+ \ref{HP-P3}h + \ref{HP-P3}i }
&=&  \frac{e_q^2 \alpha_s}{\pi^2 N_c q^2_\perp}
\int_x^1 \frac{dy_1}{y_1} \int_y^1 \frac{dy_2}{y_2} f_{\bar q^\prime } (y_2)
 \delta (\hat s (1-\xi_1)(1-\xi_2) -q^2_\perp)
\nonumber\\
  && \cdot \biggr\{  {\mathcal H}_{q\bar q 0}(\xi_1,\xi_2)
   \biggr  [ T_{q+}(-\hat y_1, x_B ) - T_{q-}(-x_B, \hat y_1 ) \biggr  ]
\nonumber\\
  &&  \  + {\mathcal H}_{\bar q q 0}(\xi_1,\xi_2) \biggr [  T_{q+}(-x_B,\hat y_1)
              -T_{q-} (-\hat y_1,x_B) \biggr ] \biggr \},
\label{Fac-HP-P3}
\end{eqnarray}
with $\hat y_1 =y_1 -x_B$ and the perturbative functions:
\begin{eqnarray}
{\mathcal H}_{q\bar q +} (\xi_1,\xi_2) &=& \frac{1-\xi_1- \xi_2}{2 N_c \xi_2 (1-\xi_2)},
 \ \ \ {\mathcal H}_{q\bar q -}(\xi_1,\xi_2) =\frac{1-\xi_1}{2 N_c \xi_2 (1-\xi_2)},
\nonumber\\
 {\mathcal H}_{\bar q q +} (\xi_1,\xi_2) &=& \frac{(1-\xi_2)^2(1-\xi_1)}{2 N_c \xi_2},
  \ \ \ {\mathcal H}_{\bar q q - } (\xi_1,\xi_2) = (1-\xi_1-\xi_2) \frac{(1-\xi_2)^2}{2 N_c \xi_2},
\nonumber\\
{\mathcal H}_{q\bar q 0}(\xi_1,\xi_2) &=& -\frac{\xi_2^2-2\xi_2 +2}{2 \xi_2^2 } (1-\xi_1-\xi_2),
\ \ \
{\mathcal H}_{\bar q q 0}(\xi_1,\xi_2) =\frac{\xi_2^2-2\xi_2 +2}{2  \xi_2^2 } (1-\xi_1).
\end{eqnarray}
$f_{\bar q^\prime} (y_2) $ is the antiquark distribution function for the flavor
which does not need to be the same as the flavor of quarks used to calculate the twist-3 matrix element $T_{q\pm}$.

\par
The studied contributions plus charge-conjugated processes give the all leading HP contributions for SSA.
All perturbative coefficient functions are at order $\alpha_s$. Combining
all possible apron flavors we obtain the factorized HP contributions as:
\begin{eqnarray}
{\mathcal W}_T\biggr\vert_{HP} &=&  \frac{\alpha_s}{\pi^2 N_c q^2_\perp }
\int_x^1 \frac{dy_1}{y_1} \int_y^{1} \frac{d y_2}{y_2}
\delta (\hat s (1-\xi_1)(1-\xi_2) -q^2_\perp)
\nonumber\\
 && \cdot \biggr  \{  {\mathcal H}_{q+} (\xi_1,\xi_2)
 \sum_{[q]}  e_q^2  f_{\bar q} (y_2) T_{q+} (y_1, x_B )
  + {\mathcal H}_{q-} (\xi_1,\xi_2)
  \sum_{ [q ]}  e_q^2  f_{\bar q}(y_2)  T_{q-} (y_1, x_B)
\nonumber\\
  && +   {\mathcal H}_{q\bar q +} (\xi_1,\xi_2)
  \sum_{[q]} e_q^2  f_{\bar q} (y_2) T_{q+}(-\hat y_1, x_B )
     + {\mathcal H}_{q \bar q -}(\xi_1,\xi_2)
     \sum_{[q]} e_q^2  f_{\bar q}T_{q-}(-\hat y_1, x_B)
\nonumber\\
  && + {\mathcal H}_{\bar q q +} (\xi_1,\xi_2)
  \sum_{[q]} e_q^2  f_{\bar q} (y_2) T_{q+}(-x_B, \hat y_1)
   + {\mathcal H}_{\bar q q -} (\xi_1,\xi_2)
   \sum_{[q]} e_q^2  f_{\bar q} (y_2) T_{q-}(-x_B, \hat y_1 )
\nonumber\\
  && +  {\mathcal H}_{q\bar q 0}(\xi_1,\xi_2) \sum_{[q,q^\prime]} e_q^2 f_{\bar q^\prime}(y_2)
   \biggr  [ T_{q+}(-\hat y_1, x_B ) - T_{q-}(-x_B,\hat y_1 ) \biggr  ]
\nonumber\\
  &&   + {\mathcal H}_{\bar q q 0}(\xi_1,\xi_2) \sum_{[q,q^\prime]} e_q^2 f_{\bar q^\prime}(y_2)\biggr [  T_{q+}(-x_B,\hat y_1)
              -T_{q-} (-y_1 +x_B,x_B) \biggr ]
\nonumber\\
 && +   {\mathcal H}_{g+}(\xi_1,\xi_2) \sum_{[q]} e_q^2 f_{g}(y_2)  T_{q+} (y_1, x_B )
     +  {\mathcal H}_{g-}(\xi_1,\xi_2) \sum_{[q]} e_q^2 f_{g}(y_2)  T_{q-} (y_1, x_B )
     \biggr \} ,
\label{HP}
\end{eqnarray}
where the notation for summing over flavors is defined as:
\begin{eqnarray}
&& \sum_{[q]} e_q^2 f_{\bar q }(y_2)  T_{q\pm}(z_1,z_2) =
\sum_{q= u,d,s,\cdots} e_q^2 \biggr [ f_{\bar q }(y_2)  T_{q\pm}(z_1,z_2)
    -f_{q }(y_2)  T_{q\mp }(-z_2,-z_1) \biggr ],
\nonumber\\
&& \sum_{[q,q^\prime ]} e_q^2 f_{\bar q^\prime }(y_2)  T_{q\pm}(z_1,z_2)
    =\sum_{q= u,d,s,\cdots, q^\prime =u,d,s,\cdots} e_q^2 \biggr [ f_{\bar q^\prime }(y_2)  T_{q\pm}(z_1,z_2)
    -f_{q^\prime }(y_2)  T_{q\mp }(-z_2,-z_1) \biggr ],
\nonumber\\
&&\sum_{[q]} e_q^2 f_{g }(y_2)  T_{q\pm}(z_1,z_2) =
\sum_{q= u,d,s,\cdots} e_q^2  f_{g}(y_2) \biggr [  T_{q\pm}(z_1,z_2)
    - T_{q\mp }(-z_2,-z_1) \biggr ].
\end{eqnarray}
It is interesting to study the limit $q^2_\perp/Q^2 \ll 1$ by using
\begin{equation}
\hat s \delta (\hat s (1-\xi_1)(1-\xi_2) -q^2_\perp)
  \approx \frac{\delta (1-\xi_1)}{(1-\xi_2)_+} + \frac{\delta (1-\xi_2)}{(1-\xi_1)_+}
      - \delta(1-\xi_1) \delta(1-\xi_2) \ln \frac{ q_\perp^2}{Q^2}.
\end{equation}
In this limit, the above contribution to
${\mathcal W}_T$  becomes:
\begin{eqnarray}
{\mathcal W}_T\biggr\vert_{HP} &=& \frac{\alpha_s}{2\pi^2 (q^2_\perp)^2}
\int_x^1 \frac{dy_1}{y_1} \int_y^{1} \frac{d y_2}{y_2}
   \cdot \biggr \{ \frac{\delta (1-\xi_2)}{(1-\xi_1)_+}
   \biggr [ \xi_1 \sum_{[q]}  e_q^2  f_{\bar q} (y_2) T_{q+} (y_1,x  )
      +\sum_{[q]}  e_q^2  f_{\bar q} (y_2) T_{q-} (y_1, x ) \biggr ]
\nonumber\\
   && +\delta (1-\xi_1) \biggr [  \frac{1+\xi_2^2}{(1-\xi_2)_+} \biggr ( 1+\frac{\xi_2-1}{N_c^2} \biggr )
    -2 \delta(1-\xi_2) \ln\frac{q^2_\perp}{Q^2} \biggr ]\sum_{[q]}  e_q^2  f_{\bar q} (y_2) T_{q+} (y_1, y_1 )
\nonumber\\
   && + \frac{ \delta(1-\xi_1)}{N_c (N_c^2-1)} (N_c^2 (1-\xi_2) -1 )((1-\xi_2)^2+\xi_2^2)
     \sum_{[q]} e_q^2 f_{g}(y_2)  T_{q+} (y_1, y_1 )
\nonumber\\
   && + \frac{\delta (1-\xi_2)}{N_c^2} \biggr [
       (1-\xi_1) \sum_{[q]} e_q^2  f_{\bar q} (y_2) T_{q-}(-y_1 +x, x )
       -\xi_1
     \sum_{[q]} e_q^2  f_{\bar q} (y_2) T_{q+}(-y_1 +x, x )  \biggr ]
 \biggr \}
\nonumber\\
    &&  \cdot   \left [ 1 + {\mathcal O}(q^2_\perp/Q^2) \right ].
\label{qtlim}
\end{eqnarray}
It is noted that in the limit SGP contributions appear.
If we take the limit $q^2_\perp/Q^2 \ll 1$
in the tree-level results for the parotnic ${\mathcal W}_T$'s in Eq.(\ref{qWT}) and Eq.(\ref{gWT})
instead of  the factorized results in Eq.(\ref{Fac-Fig2}) and Eq.(\ref{qHP}), we will not obtain
the SGP contributions. However, the SGP contributions can be derived by using parotnic ${\mathcal W}_T$s
in the limit beyond tree-level\cite{MS4}.

\par
The factorized results of Fig.2 and
Fig.3 have been derived in \cite{JQVY1} with the method of the diagram expansion mentioned
in the Introduction.
By rewriting the above results with partonic variables
which are defined as:
\begin{eqnarray}
 \hat t &=& (y_1 P_A-q)^2 = -\hat s \xi_2(1-\xi_1) -q^2_\perp, \ \ \ \ \hat s = y_1 y_2 s,
\nonumber\\
 \hat u &=& (y_2 P_B -q)^2 = -\hat s \xi_1 (1-\xi_2) -q^2_\perp,  \ \ \  Q^2 =q^2 = \xi_1\xi_2 \hat s -q^2_\perp,
\end{eqnarray}
we find that our results agree with them in \cite{JQVY1}. Recently, the results corresponding
to the contributions from Fig.4 have been derived with the method of the diagram expansion
in \cite{KKnew}. Again our results in Eq. (\ref{Fac-HP-P3}) agree with those in \cite{KKnew}.

\par\vskip20pt
\noindent
{\bf 5. Soft-Gluon-Pole Contributions}
\par\vskip10pt
The SGP contributions comes from the case when one gluon with zero momentum
enters hard scattering. They may come from the $qg$-, $q\bar q$-
and the $gg$-contributions.
The $qg$- and $q\bar q$ contributions are factorized with the quark-gluon correlator $T_{q+}(x,x)=T_{q-}(x,x)$.
Later we will show that the $q\bar q$-contributions need not to be studied, because
it is automatically included in the factorized form obtained from
the $qg$-contributions.
The $gg$-contributions are  factorized with the purely gluonic correlator defined in Eq.(\ref{SGP-GG}).
We will discuss these two types of contributions in this section separately.

\begin{figure}[hbt]
\begin{center}
\includegraphics[width=4cm]{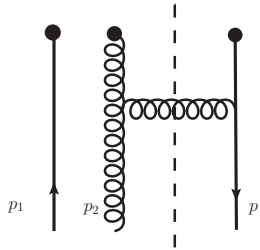}
\end{center}
\caption{The diagram for $T_{q\pm}(x,x)$ at one-loop. The black dots denote the insertion of operators
used to define $T_{q\pm}(x_1,x_2)$ in Eq.(\ref{def-tw3q}). }
\label{SGP-Tq}
\end{figure}
\par

\par\vskip10pt
\noindent
{\bf 5.1. The $qg$-Contributions}
\par\vskip10pt
We have given the results of the $qg$-contributions for $T_{q\pm}(x_1,x_2)$ at tree-level in Eq.(\ref{tree-Tq}). At this order
one simply has $T_{q\pm}(x,x)=0$. However,
beyond the tree-level, $T_{q\pm}(x,x)$ can be nonzero. As found in \cite{MS3,MS4}, at one-loop level
there is only one diagram giving nonzero contribution to $T_{q\pm}(x,x)$ in the light-cone-
or Feynman gauge.
The calculation of the diagram is straightforward. The contribution has an U.V.- and a collinear divergence. Both
are regularized with the dimensional regularization as poles of $\epsilon=4-d$. After extracting
the U.V. pole  we have\cite{MS3,MS4}:
\begin{equation}
T_{q\pm} (x,x, \mu ) =-  {\mathcal C}^{qg} \frac{ g_s \alpha_s}{4} N_c (N_c^2-1) x_0 \sqrt{2x_0} \delta (x_0-x)
 \left [ \left (-\frac{2}{\epsilon_c} \right ) + \ln\frac{e^\gamma \mu^2}{4\pi \mu_c^2} \right ]
 + {\mathcal O}(g_s\alpha_s^2) ,
\label{Tq-SGP}
\end{equation}
where the pole is the collinear divergence with the index $c$. $\mu$ is the renormaliation scale related
to the U.V. pole, and $\mu_c$ is that related to the collinear pole.
\par
\begin{figure}[hbt]
\begin{center}
\includegraphics[width=14cm]{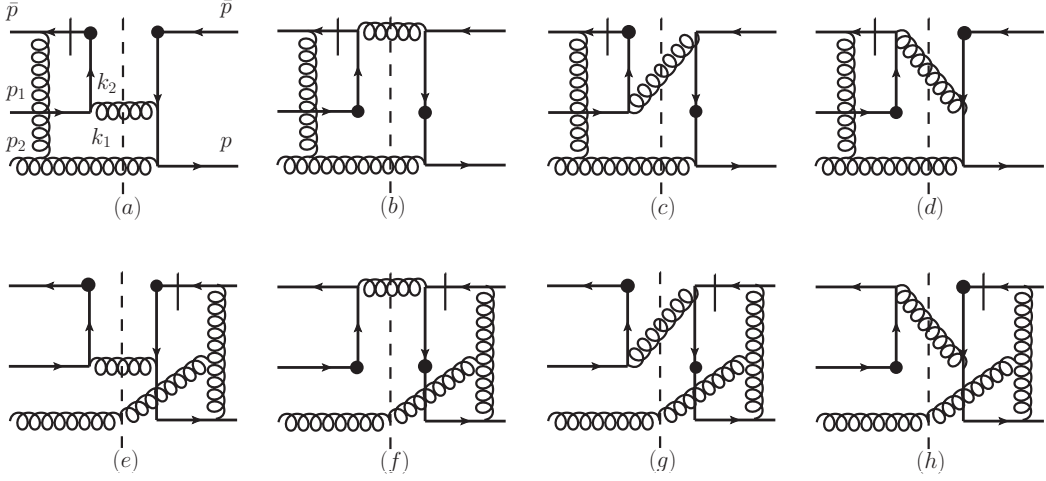}
\end{center}
\caption{The diagrams for the amplitude $\bar q + [ q+ G ] \to \gamma^* +X \to \bar q + q$ for SGP contributions. The black dots represent the insertion of electromagnetic current operator.  }
\label{SGP-P1}
\end{figure}

To find out the SGP contributions it is convenient to work with the light-cone gauge $n\cdot G=0$.
We consider
a special class of diagrams which represent a part of one-loop corrections
to those given in Fig.2. These diagrams are obtained from Fig.2. by adding a gluon. They are given in 
Fig.\ref{SGP-P1}.
In the first four diagrams the gluon is emitted by the initial gluon and is absorbed by the final quark.
In the last four diagrams the initial gluon goes across the cut represented
by the broken line and emits a virtual gluon absorbed by the outgoing quark.
\par
The contributions from Fig.\ref{SGP-P1} contain a collinear divergence. In the first 
four diagrams, the divergence appears when the lowest gluon crossing the cut is collinear
to the $+$-direction. In the last four diagrams it appears when 
the gluon emitted by the outgoing quark is collinear. 
Because the contributions from Fig.\ref{SGP-P1} are one-loop corrections to Fig.2,
one may expect that the collinearly divergent parts of the contributions
can be re-produced in the factorized form of the contributions from Fig.2. in Eq.(\ref{Fac-Fig2}),
where one replaces $T_{q\pm}(x_1,x_2)$ with the corresponding one-loop $T_{q\pm}(x_1,x_2)$.  
As discussed in detail in \cite{MS4}, this is not the case, because the color factor
here does not match. Even if one neglects the color factor, the divergences still can not be re-produced.  

\par

\par
Analyzing the collinear divergences
in the contributions of Fig.\ref{SGP-P1}, one finds that
the collinear divergences should be factorized with $T_{q+}(x,x)=T_{q-}(x,x)$.
Taking Fig.\ref{SGP-P1}a as an example, the added gluon is with momentum
$k_1$. If $k_1$ is collinear to the $+$-direction, i.e., $k_1^\mu \sim  (1,\lambda^2,\lambda,\lambda)$
with $\lambda \ll 1$, one can find that the gluon exchanged between the initial gluon
and the initial antiquark is soft with the on-shell condition of the cut propagator.
In fact, this gluon is a Glauber gluon with the momentum
$\sim (\lambda^2,\lambda^2,\lambda,\lambda)$.
Comparing Fig.\ref{SGP-P1}a with Fig.5, one can identify that the gluon crossing the cut in Fig.5
corresponds to the collinear gluon with $k_1$ in Fig.\ref{SGP-P1}a.
If the collinear gluon is contained
in $T_{q\pm}$, the Glauber gluon should be taken as the gluon entering hard scattering.
Since it is a Glauber gluon with vanishing momentum, the divergent parts of Fig.\ref{SGP-P1}
should be factorized with $T_{q+}(x,x)$. This is the reason why the SGP
contributions appear.
\par
Performing the same analysis for Fig.\ref{SGP-P1}b,  Fig.\ref{SGP-P1}c and Fig.\ref{SGP-P1}d in the case that the gluon crossing the cut is collinear, one will find that the gluon exchanged between the initial antiquark and the
initial gluon is a Glauber gluon. For the last four diagrams the gluon emitted by the outgoing antiquark
in the right part is a Glauber gluon, if the gluon emitted by the outgoing quark is collinear.
Therefore, the collinear divergences in these diagrams are related to the Glauber gluon.
It should be noted that only the diagrams in Fig.\ref{SGP-P1} contain
such a collinear divergence related to a Glauber gluon.
\par
Before giving the results, the following facts should be pointed out.
In Feynman gauge, one has to consider
more diagrams which contain the collinear divergence, e.g., instead of that the collinear gluon is attached
to the initial gluon in the left part of Fig.\ref{SGP-P1}, the gluon can also be attached
to the initial antiquark. Such diagrams are finite in the light-cone gauge,
at least for most cases studied here with an exception which will be discussed in Sect. 6.
In the following we will work in the light-cone gauge $n\cdot G =0$.

\par
The contributions of Fig.\ref{SGP-P1} contain an integration of a loop-momentum.
It is easy to find the collinearly divergent part of the contributions
by expanding the integrand in $\lambda$, where the collinear gluon has the momentum
$\sim (1,\lambda^2, \lambda,\lambda)$. We find the collinearly divergent part of the contributions from Fig.\ref{SGP-P1} as:
\begin{eqnarray}
{\mathcal W}_T\biggr\vert_{Fig.\ref{SGP-P1}} &=& {\mathcal C}^{qg}
\frac{ e_q^2 g_s\alpha_s^2}{2\pi^2} \frac{N_c^2-1}{2 N_c}
   \frac{\sqrt{2x_0}}{ x_0}
   \left ( -\frac{2}{\epsilon_c} \right )
 \cdot \left [ \frac{x^2 -2x x_0 -x_0^2 y} {(x_0-x)^2 (1-y) s} \delta (u)
       + \frac{x^2 +x_0^2 y^2}{(x_0-x)(1-y)} \delta '(u) \right ],
\nonumber\\
    && \delta (u) =\delta (s(x_0-x)(1-y)-q^2_\perp).
\label{WFig6}
\end{eqnarray}
In the above the pole in $\epsilon_c=4-d$ represents the collinear divergence.
The $\delta$-function from the on-shell condition of the intermediate gluon exchanged between 
quarks also depends on the loop momentum and needs to be expanded in $\lambda$. 
This results in the terms with the derivative of the $\delta$-function. 
The last four diagrams do not contain terms with
the derivative of the $\delta$-function. 
With the result of $T_{q+}(x,x)$ from the $qg$-contribution in Eq.(\ref{Tq-SGP})
we can derive the factorized form:
\begin{eqnarray}
{\mathcal W}_T\biggr\vert_{Fig.\ref{SGP-P1}} &=&  \frac{ e_q^2 \alpha_s} {\pi^2 N_c q^2_\perp}
  \int_x^1 \frac{dy_1}{y_1} \int_y^1 \frac{d y_2}{y_2} f_{\bar q} (y_2)
   \delta (\hat s( 1-\xi_1)(1-\xi_2)-q^2_\perp)
\nonumber\\
   && \ \ \ \  \cdot \biggr  [  \tilde {\mathcal S}_{Gq} (\xi_1,\xi_2)
   \left ( y_1 \frac{\partial T_{q+}(y_1,y_1)}{\partial y_1} \right )
      + {\mathcal S}_{Gq} (\xi_1,\xi_2) T_{q+}(y_1,y_1) \biggr  ],
\nonumber\\
    \tilde{\mathcal S}_{Gq} (\xi_1,\xi_2)  &= &  \frac{\xi_1^2 +\xi_2^2}{N_c (1-\xi_2)}, \ \ \ \
    {\mathcal S}_{Gq} (\xi_1,\xi_2) = \frac{2\xi_1(1-\xi_1)^2-\xi_2^3-\xi_1\xi_2(2-\xi_1)}
        { N_c (1-\xi_1)(1-\xi_2)}.
\end{eqnarray}
We note that the perturbative coefficeint  function here is at the same order of $\alpha_s$ as
those of HP contributions because $T_{q+}(y_1,y_1)$ is at the order of $g_s\alpha_s$.
\par
\begin{figure}[hbt]
\begin{center}
\includegraphics[width=14cm]{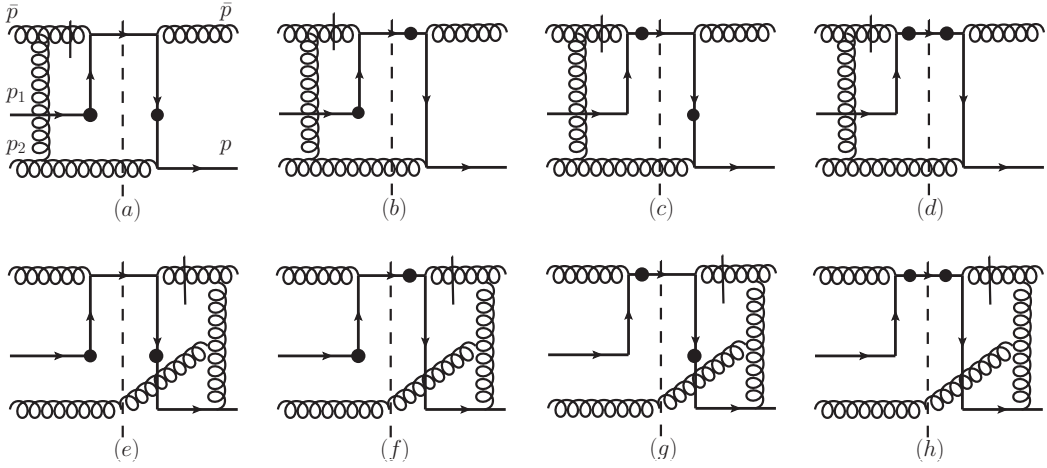}
\end{center}
\caption{The diagrams for the amplitude $g + [ q+ G ] \to \gamma^* +X \to g+ q$ for SGP contributions. The
black dots represent the insertion of electromagnetic current operator.  }
\label{SGP-P2}
\end{figure}
\par
If we replace $h_B$ with a gluon, one obtains similar diagrams from
the SGP contributions from the $qg$-contributions. These diagrams are given in Fig.\ref{SGP-P2}.
The collinearly divergent part of the contributions belong
to the SGP contributions. We have calculated the collinear divergences in these diagrams
in the light-cone gauge and in Feynman gauge. The same results are obtained. This corresponds
to the situation with $T_{q\pm}(x,x)$ with Fig.\ref{SGP-Tq}, only the same one diagrams
in the two gauges
gives the result in Eq.(\ref{Tq-SGP}). From Fig.\ref{SGP-P2} we have:
\begin{eqnarray}
{\mathcal W}_T\biggr\vert_{Fig.\ref{SGP-P2}}
    &=& {\mathcal C}^{qg} \frac{ e_q^2 g_s\alpha_s^2 N_c^2}{4 \pi^2 } \sqrt{2x_0}
      \biggr [ -\frac{s (1-y)(x^2+2x x_0 (y-2) + 2x_0^2 (y^2-2y+2)) }{q^2_\perp }\delta '(u)
\nonumber\\
    &&  + \frac{s(1-y)\delta(u)}{x_0 (q^2_\perp)^2}
      \biggr (
       (x_0-x) ( -x y + 2 x +3x_0y-4x_0)
   - x_0^2 y (y^2 +(1-y)^2)      \biggr ) \biggr ]\left (-\frac{2}{\epsilon_c } \right ).
\nonumber\\
\end{eqnarray}
With the result of $T_{q+}(x,x)$ from the $qg$-contribution in Eq.(\ref{Tq-SGP})
we can derive the factorized form:
\begin{eqnarray}
{\mathcal W}_T\biggr\vert_{Fig.\ref{SGP-P2}}  &=& \frac{ e_q^2 \alpha_s }{\pi^2 N_c  q^2_\perp}
 \int_x^1 \frac{dy_1}{y_1} \int_y^1\frac{dy_2}{y_2}
 f_g (y_2) \delta(\hat s(1-\xi_1)(1-\xi_2)-q^2_\perp)
\nonumber\\
  && \cdot  \biggr [  \tilde{\mathcal S}_{Gg} (\xi_1,\xi_2)
   \left ( y_1\frac{\partial T_{q+}(y_1,y_1)  }{\partial y_1} \right )
   + {\mathcal S}_{Gg}(\xi_1,\xi_2) T_{q+} (y_1,y_1)
 \biggr ],
\nonumber\\
    \tilde{\mathcal S}_{Gg} (\xi_1,\xi_2) &=& - \frac{N_c^2}{N_c^2-1}
     \left ( \xi_1^2 + 2\xi_1 (\xi_2-2) + 2 (\xi_2^2 -2\xi_2 +2) \right ),
\nonumber\\
       {\mathcal S}_{Gg} (\xi_1,\xi_2) &=& - \frac{N_c^2}{N_c^2-1}
          \biggr (
  - 3 \xi_1\xi_2 + 6\xi_1 -2 \xi_1^2 + 3\xi_2 -4  -\frac{\xi_2 (\xi_2^2 +(1-\xi_2)^2)}{1-\xi_1} \biggr ).
\label{Fac-SGP-P2}
\end{eqnarray}
The factorized results have also been also derived with the method of diagram expansion in \cite{JQVY1}.

\par
\begin{figure}[hbt]
\begin{center}
\includegraphics[width=7cm]{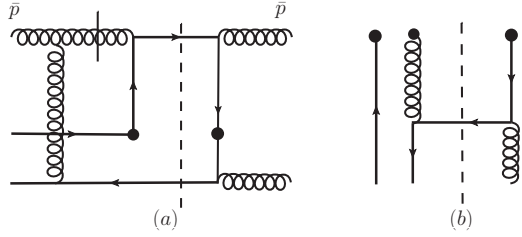}
\end{center}
\caption{(a). The possible SGP-contributions from
the $q\bar q$-contributions. (b). The diagram for $T_{q\pm}(x,x)$
in the gauge $n\cdot G=0$ from the $q\bar q$-contributions. See the discussion in text.  }
\label{SGP-P3}
\end{figure}
\par
\par
In the case when $h_B$ is replaced by a gluon, one can have
the SGP contribution from the $q\bar q$-contributions. An typical diagram is given in Fig.\ref{SGP-P3}.
One can also obtain $T_{q\pm}(x,x)$ from the $q\bar q$-contributions at this order.
The diagram for it is given by Fig.\ref{SGP-P3}b. It is easy to find
that the SGP contribution is included in the factorized form in Eq.(\ref{Fac-SGP-P2}).
\par
Combining contributions of all flavors the SGP contributions can be factorized with the quark-gluon
twist-3 matrix element as:
\begin{eqnarray}
{\mathcal W}_T\biggr\vert_{SGPF} &=&  \frac{ \alpha_s} {\pi^2 N_c q^2_\perp}
  \int_x^1 \frac{dy_1}{y_1} \int_y^1 \frac{d y_2}{y_2}
   \delta (\hat s( 1-\xi_1)(1-\xi_2)-q^2_\perp)
\nonumber\\
   &&   \cdot \biggr  [  \tilde {\mathcal S}_{Gq} (\xi_1,\xi_2)
   \sum_{[q]} e_q^2 f_{\bar q} (y_2) \left ( y_1 \frac{\partial T_{q+}(y_1,y_1)}{\partial y_1} \right )
      + {\mathcal S}_{Gq} (\xi_1,\xi_2) \sum_{[q]} e_q^2 f_{\bar q} (y_2) T_{q+}(y_1,y_1)
\nonumber\\
  && +  \tilde{\mathcal S}_{Gg} (\xi_1,\xi_2)
   \sum_{ [ q]} e_q^2 f_g (y_2) \left ( y_1\frac{\partial T_{q+}(y_1,y_1)  }{\partial y_1} \right )
   + {\mathcal S}_{Gg}(\xi_1,\xi_2)  \sum_{[q]} e_q^2 f_g (y_2) T_{q+} (y_1,y_1)
      \biggr  ].
\end{eqnarray}
The above results agree with those in \cite{JQVY1,KKnew} derived with other method.
Again, in the case of $q_\perp^2 /Q^2 \ll 1$ this contribution takes a simplified form:
\begin{eqnarray}
{\mathcal W}_T\biggr\vert_{SGPF} &=&  \frac{ \alpha_s} { \pi^2 N_c^2 (q^2_\perp)^2}
  \int_x^1 \frac{dy_1}{y_1} \int_y^1 \frac{d y_2}{y_2}
  \cdot \biggr  [  (1+\xi_1^2) \delta (1-\xi_2)
   \sum_{[q]} e_q^2 f_{\bar q} (y_2) \left ( y_1 \frac{\partial T_{q+}(y_1,y_1)}{\partial y_1} \right )
\nonumber\\
     &&  +  \left ( \frac{\delta (1-\xi_2}{(1-\xi_1)_+} ( 2\xi_1^3 -3\xi_1^2 -1)
          -\frac{\delta (1-\xi_1)}{(1-\xi_2)_+} \xi_2 (1+\xi_2^2)
            +2 \delta (1-\xi_1)\delta (1-\xi_2) \ln\frac{q^2_\perp}{Q^2} \right )
\nonumber\\
     && \cdot  \sum_{[q]} e_q^2 f_{\bar q} (y_2) T_{q+}(y_1,y_1)  + \frac{ N_c^3 \xi_2}{N_c^2-1}
         (\xi_2^2 + (1-\xi_2)^2) \delta (1-\xi_1)  \sum_{[q]} e_q^2 f_g (y_2) T_{q+}(y_1,y_1)
      \biggr  ]
\nonumber\\
    && + \cdots
\end{eqnarray}
where $\cdots$ stand for contributions  suppressed by $q^2_\perp/Q^2$.

\par\vskip10pt
\noindent
{\bf 5.2. The $gg$-Contributions}
\par\vskip10pt
\par
At the order we consider, there is no HP contribution from the $gg$-contributions. But, it is possible
that there are leading SGP contributions from ${\mathcal W}_T$ at one-loop level, similar to cases considered
in the above.  We consider first the gluonic twist-3 matrix elements
in Eq.(\ref{GF-def}). These functions are zero at tree-level.

\begin{figure}[hbt]
\begin{center}
\includegraphics[width=8cm]{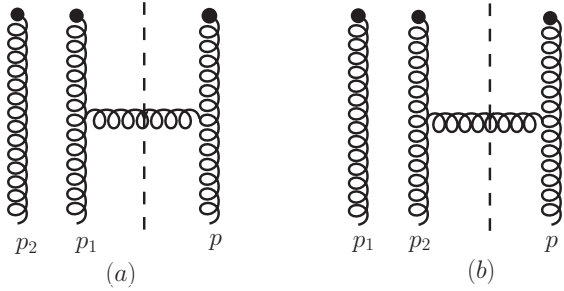}
\end{center}
\caption{The diagrams for $G_{d1,f1}(x)$ and $G_{d2,f2}(x)$ in the light-cone gauge.   }
\label{GGG}
\end{figure}
\par
At one-loop level, the functions become nonzero. They receive nonzero contributions from the diagrams
given in Fig.\ref{GGG} in the light-cone gauge. In Feynman gauge there are more diagrams. In this subsection
we will work with the light-cone gauge.
For the factorization studied below we only need to calculate
Fig.\ref{GGG}a and the corresponding diagrams for ${\mathcal W}_T$. The contributions from Fig.\ref{GGG}b 
and the corresponding contributions to ${\mathcal W}_T$ can be obtained from the permutation 
of the two initial gluons.  We will only give results
from Fig.\ref{GGG}a and the corresponding results of ${\mathcal W}_T$. 
We obtain:
\begin{eqnarray}
G_{d1} (x) &=& -\frac{g_s \alpha_s}{2 \sqrt{2} } (N_c^2-4)(N_c^2-1)
\delta (x-\bar x_0)  \left [ \left (-\frac{2}{\epsilon_c} \right ) + \ln\frac{e^\gamma \mu^2}{4\pi \mu_c^2} \right ] d_1,
\nonumber\\
G_{d2}  (x) &=& \frac{g_s \alpha_s}{4 \sqrt{2} } (N_c^2-4)(N_c^2-1)
 \delta (x-\bar x_0)  \left [ \left (-\frac{2}{\epsilon_c} \right ) + \ln\frac{e^\gamma \mu^2}{4\pi \mu_c^2} \right ]   d_2,
\nonumber\\
G_{f1} (x) &=& \frac{g_s \alpha_s}{2 \sqrt{2} } N_c^2(N_c^2-1)
 \delta (x-\bar x_0)  \left [ \left (-\frac{2}{\epsilon_c} \right ) + \ln\frac{e^\gamma \mu^2}{4\pi \mu_c^2} \right ]f_1,
\nonumber\\
G_{f2}  (x) &=& -\frac{g_s \alpha_s}{4 \sqrt{2} } N_c^2(N_c^2-1)
\delta (x-\bar x_0)\left [ \left (-\frac{2}{\epsilon_c} \right ) + \ln\frac{e^\gamma \mu^2}{4\pi \mu_c^2} \right ]f_2,
\label{Result-GF}
\end{eqnarray}
with the parameters $d_{1,2}$ and $f_{1,2}$ related to ${\mathcal F}_\pm^{gg}$ in Eq.(\ref{ndmggF})
and ${\mathcal D}_\pm^{gg}$ as:
\begin{eqnarray}
 && d_1 =(1-x_0) \left (
      x_0 {\mathcal D}_+^{gg}   + {\mathcal D}_-^{gg}    \right ), \ \ \
       d_2 = {\mathcal D}_+^{gg}   + {\mathcal D}_-^{gg} ,
\nonumber\\
 &&  f_1 =(1-x_0) \left (
      x_0 {\mathcal F}_+^{gg}   + {\mathcal F}_-^{gg}  \right ), \ \ \
      f_2 ={\mathcal D}_+^{gg}   + {\mathcal D}_-^{gg} .
\end{eqnarray}

\begin{figure}[hbt]
\begin{center}
\includegraphics[width=14cm]{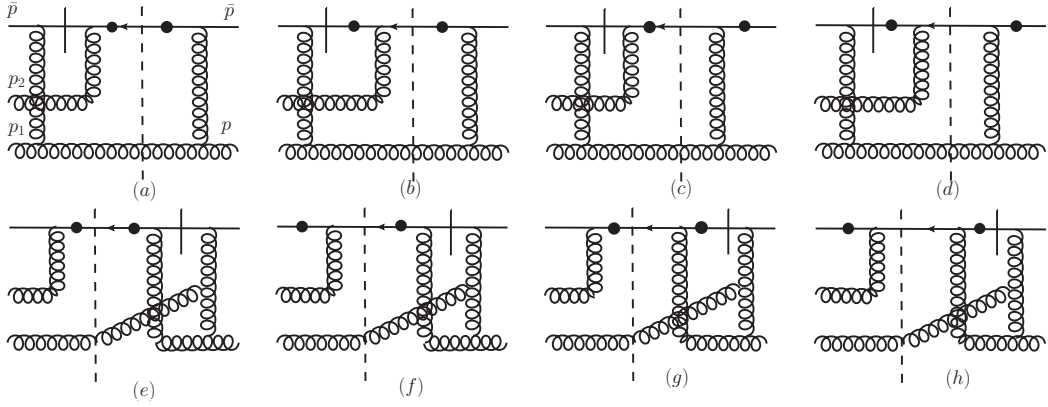}
\end{center}
\caption{The diagrams for the amplitude
$\bar q + (G+G) \to \gamma^*  + X \to \bar q + G $ at one-loop for possible SGP contributions.  }
\label{SGP-GGG}
\end{figure}
\par
\par
The corresponding contributions to ${\mathcal W}_T$ are given by diagrams
in Fig.\ref{SGP-GGG}. The results for the color antisymmetric gluon state are:
\begin{eqnarray}
{\mathcal W}_T\biggr\vert_{Fig.\ref{SGP-GGG}}   &= & -\frac{e_q^2 g_s\alpha_s^2} {4\sqrt{2}\pi^2 N_c}
  \frac{N_c^2 (N_c^2-1)}{\bar x_0^3 (1-y) } \left ( -\frac{2}{\epsilon_c} \right )
     \biggr \{ \delta' (u)  \biggr [ 2 f_1 \left ( 2 x^2 + 2 x\bar x_0 (y-2) +\bar x_0^2 (y-2)^2 \right )
\nonumber\\
  && - f_2 \left ( 4x^2 +4 x \bar x_0 (y-2) +\bar x_0^2 (y^2-6y+6) \right ) \biggr ]
\nonumber\\
  && - \frac{\delta (u)}{s  } \biggr [4 f_1 \left ( 2x +\bar x_0 (y-2) \right )
       + 2 f_2 \frac{x(5y-4)+\bar x_0 (3y^2 -7y +4)}{1-y} \biggr ]  \biggr \},
\nonumber\\
  u &=&s(\bar x_0 -x)(1-y) -q^2_\perp.
\end{eqnarray}
Replacing the color factor $N_c^2 (N_c^2-1)$ with $-(N_c^2-4)(N_c^2-1)$ and $f_{1,2}$ with $d_{1,2}$,
respectively, one obtains ${\mathcal W}_T$ from Fig.\ref{SGP-GGG} with the color structure of $d^{abc}$.
With the results of the gluonic twist-3 matrix elements in Eq.(\ref{Result-GF}) we can derive
the factorized form from the SGP contribution from Fig.\ref{SGP-GGG} by combining
all flavors as:
\begin{eqnarray}
{\mathcal W}_T\biggr\vert_{SGPG} &=& \frac{\alpha_s} {\pi^2 N_c q_\perp^2}
\int_x^1 \frac{dy_1}{y_1} \int_y^1 \frac{d y_2}{y_2}
   \delta (\hat s( 1-\xi_1)(1-\xi_2)-q^2_\perp)
\nonumber\\
  && \cdot \biggr \{ \sum_{i=1,2} \tilde{\mathcal S}_{Gi} (\xi_1,\xi_2) \sum_{q} e_q^2 \biggr [ f_{\bar q} (y_2)
  \left ( y_1 \frac{\partial G_{fi}(y_1,y_1)} {\partial y_1}
        + y_1 \frac{\partial G_{di} (y_1,y_1)} {\partial y_1} \right )
\nonumber\\
  &&         + f_{q}(y_2)
  \left ( y_1 \frac{\partial G_{fi}(y_1,y_1)} {\partial y_1}
        - y_1 \frac{\partial G_{di}(y_1,y_1 )} {\partial y_1} \right )  \biggr ]
\nonumber\\
   &&  + \sum_{i=1,2} {\mathcal S}_{Gi} (\xi_1,\xi_2) \sum_{q} e_q^2 \biggr [ f_{\bar q} (y_2)
  \left ( G_{fi}(y_1,y_1)
         + G_{di}(y_1,y_1) \right )
\nonumber\\
  &&         + f_{q}(y_2)
  \left ( G_{fi}(y_1,y_1)
        - G_{di}(y_1,y_1) \right )  \biggr ]  \biggr \}
\end{eqnarray}
with the pertubative functions:
\begin{eqnarray}
\tilde{\mathcal S}_{G1}(\xi_1,\xi_2) &=&  \frac{1-\xi_1}{1-\xi_2}
\left ( 2\xi_1^2 +2 \xi_1 (\xi_2-2)  +(\xi_2-2)^2 \right ) ,
\nonumber\\
\tilde{\mathcal S}_{G2}(\xi_1,\xi_2) &=& \frac{1-\xi_1}{1-\xi_2}
  \left ( 4\xi_1^2 +4 \xi_1 (\xi_2-2) +\xi_2^2-6\xi_2 +6 \right ),
\nonumber\\
{\mathcal S}_{G1} (\xi_1,\xi_2) &=& -\frac{1-\xi_1}{1-\xi_2} \left ( 6\xi_1^2+ 4 \xi_1 (2\xi_2-3)
     +3\xi_2^2 -10\xi_2 + 8  \right ),
\nonumber\\
 {\mathcal S}_{G2} (\xi_1,\xi_2) &=& -\frac{1-\xi_1}{1-\xi_2} \left ( 12\xi_1^2 +6 \xi_1 (3\xi_2-4)
     +7\xi_2^2 -20\xi_2 +14 \right ) .
\end{eqnarray}
From the above results we can derive the result in the limit $q_\perp \to 0$ as:
\begin{eqnarray}
 {\mathcal W}_T\biggr\vert_{SGPG} &=& \frac{\alpha_s} {\pi^2 N_c (q_\perp^2)^2}
\int_x^1 \frac{dy_1}{y_1} \int_y^1 \frac{d y_2}{y_2} (1-\xi_1)\delta (1-\xi_2)
\nonumber\\
  && \cdot \biggr \{ \sum_{i=1,2} \tilde{\mathcal S}_{\perp i} (\xi_1,\xi_2) \sum_{q} e_q^2 \biggr [ f_{\bar q} (y_2)
  \left ( y_1 \frac{\partial G_{fi}(y_1,y_1)} {\partial y_1}
        + y_1 \frac{\partial G_{di} (y_1,y_1)} {\partial y_1} \right )
\nonumber\\
  &&         + f_{q}(y_2)
  \left ( y_1 \frac{\partial G_{fi}(y_1,y_1)} {\partial y_1}
        - y_1 \frac{\partial G_{di}(y_1,y_1 )} {\partial y_1} \right )  \biggr ]
\nonumber\\
   &&  + \sum_{i=1,2} {\mathcal S}_{\perp i} (\xi_1,\xi_2) \sum_{q} e_q^2 \biggr [ f_{\bar q} (y_2)
  \left ( G_{fi}(y_1,y_1)
         + G_{di}(y_1,y_1) \right )
\nonumber\\
  &&         + f_{q}(y_2)
  \left ( G_{fi}(y_1,y_1)
        - G_{di}(y_1,y_1) \right )  \biggr ]  \biggr \}
\end{eqnarray}
with:
\begin{eqnarray}
\tilde{\mathcal S}_{\perp 1} &=& 2\xi_1^2 -2\xi_1 +1, \ \ \ \ \ \ \ \
\tilde{\mathcal S}_{\perp 2} = 4\xi_1^2-4\xi_1 + 1,
\nonumber\\
{\mathcal S}_{\perp 1} &=& -\left (6\xi_1^2 -4\xi_1^2 +1 \right ),  \ \ \ \
{\mathcal S}_{\perp 2} = -\left (12\xi_1^2 - 6\xi_1^2 +1 \right ).
\end{eqnarray}
The above the SGP contributions are leading contributions in the limit.

\par\vskip20pt
\noindent
{\bf 6. SQP-Contributions}
\par\vskip10pt

\par
\begin{figure}[hbt]
\begin{center}
\includegraphics[width=11cm]{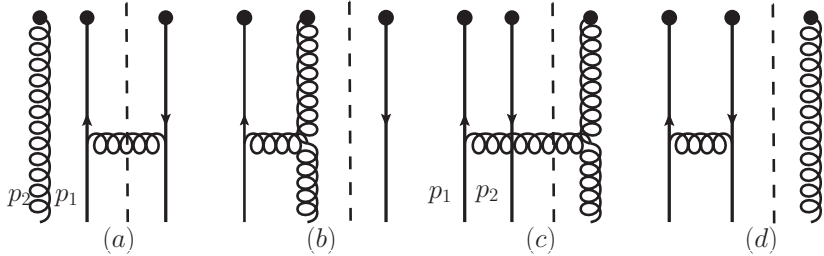}
\end{center}
\caption{The diagrams for the twist-3 matrix elements with $x_2=0$ in the gauge
$n\cdot G=0 $. The first two diagrams are for $x_1>0$
in $qg$-contributions. The later two diagrams are for $x_1<0$ in $q\bar q$-contributions.}
\label{SFP-T3R}
\end{figure}

\par
Similarly to the twist-3 matrix elements for SGP contributions, the twist-3 matrix elements for SQP contributions
are zero at tree-level, because one can not define a quark state with zero momentum. Beyond
tree-level, they can be nonzero.
In the light-cone gauge $n\cdot G =0$, one can find two possible diagrams at one-loop
for the $qg$-contributions and the $q\bar q$-contributions. They are given in Fig.\ref{SFP-T3R}.
It is easy to find that Fig.\ref{SFP-T3R}b and Fig.\ref{SFP-T3R}d will give zero contribution.
We have for the $qg$-contributions from Fig.\ref{SFP-T3R}a as:
\begin{eqnarray}
T_{q+}(x,0)  &=& {\mathcal C}^{qg}  g_s \alpha_s \frac{N_c^2-1}{4 N_c}
\frac{x\sqrt{2x_0}}{x_0} \delta (x-\bar x_0 ) \left [  -\left (\frac{2}{\epsilon_c}\right )
  + \ln\frac{ e^\gamma \mu^2}{ 4\pi \mu_c^2} \right ],
\nonumber\\
T_{q- }(x,0)  &=& 0.
\label{SQP-T3P1}
\end{eqnarray}
We have for the $q\bar q$-contributions from Fig.\ref{SFP-T3R}c as:
\begin{eqnarray}
T_{q+} (x,0) &=&  \left ( {\mathcal C}_+^{q\bar q} - {\mathcal C}_-^{q\bar q} \right ) g_s \alpha_s \delta(x+\bar x_0) \frac{N_c (N_c^2-1)}{4 }
 \frac{\sqrt{2x_0\bar x_0}}{x_0}
 \left [  -\left (\frac{2}{\epsilon_c} \right )
  + \ln\frac{ e^ \gamma \mu^2}{ 4\pi \mu_c^2} \right ] ,
\nonumber\\
T_{q-} (x,0) &=& \left ( {\mathcal C}_+^{q\bar q} +  {\mathcal C}_-^{q\bar q} \right )
    g_s \alpha_s \delta(x+\bar x_0) \frac{N_c (N_c^2-1)}{4 }
 \frac{\sqrt{2x_0\bar x_0}}{x_0} \bar x_0^2
  \left [  -\left (\frac{2}{\epsilon_c} \right )
  + \ln\frac{ e^ \gamma \mu^2}{ 4\pi \mu_c^2} \right ].
\label{SQP-T3P2}
\end{eqnarray}
It is noted that in the above  $x$ is negative. It implies that an antiquark with the momentum fraction $-x$ enters a hard scattering.
\par
\begin{figure}
\begin{center}
\includegraphics[width=14cm]{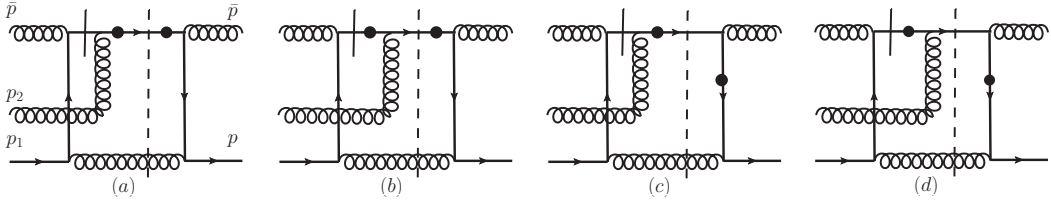}
\end{center}
\caption{The diagrams in the $n\cdot G=0$ gauge which give the soft fermion pole contributions to SSA. }
\label{SFP-P1}
\end{figure}
\par\par
The SQP contributions from the $qg$-contributions to ${\mathcal W}_T$ are given by diagrams
in Fig.\ref{SFP-P1} in the gauge $n\cdot G=0$. Following the analysis similar to that of Fig.\ref{SGP-P1},
one can see that the vertical quark line in the left part of diagrams carries
the momentum $k_q$ at the order of $k_q^{\mu} \sim (\lambda^2,\lambda^2,\lambda,\lambda)$,
if the gluon at the bottom
crossing the cut is collinear, i.e., its momentum scales like $(1,\lambda^2,\lambda,\lambda)$.
Factorizing the collinear gluon into the corresponding twist-3 matrix elements, one can realize that
in the left part of diagrams, there is a gluon combined with a soft quark entering the hard scattering.
Therefore, the collinearly divergent contributions are SQP-contributions.
\par

\par
It is straightforward to find the divergent contributions from Fig.\ref{SFP-P1}:
\begin{eqnarray}
{\mathcal W}_T\biggr\vert_{Fig.\ref{SFP-P1}}  &=&  {\mathcal C}^{qg} \frac{ e_q^2 g_s\alpha_s^2}
  {16\pi^2 N_c^2}
     \frac{ \sqrt{2x_0} \delta ( s(\bar x_0 -x)(1-y) -q^2_\perp)}{q^2_\perp (1-x_0) }
   \biggr [ \frac{ ( x(2x-3(x_0-1)(y-2)) }{1-x_0}
\nonumber\\
    && \ \ \  +(1-x_0) ( y^2 -5 y +5)
      -\vert \lambda_q \vert (x(y-2)-(x_0-1) (y^2-3y +3) )
     \biggr ]\left (-\frac{2}{\epsilon_c } \right ).
\label{WTSFP}
\end{eqnarray}
Again the quark-spin independent part should be factorized with the combination $T_{+q}(x,0) + T_{-q}(x,0)$,
and the contribution with $\vert \lambda_q \vert$ should be factorized with $T_{+q}(x,0)-T_{-q}(x,0)$.
With the results in Eq.(\ref{SQP-T3P1}) we have:
\begin{eqnarray}
{\mathcal W}_T\biggr\vert_{Fig.\ref{SFP-P1}}  &=& \frac{e_q^2\alpha_s }{ 2\pi^2 N_c  q_\perp^2}
 \int_x^1 \frac{dy_1}{y_1} \int_y^1\frac{dy_2}{y_2}
 f_g (y_2)  \delta(\hat s (1-\xi_1)(1-\xi_2)-q^2_\perp)
\nonumber\\
   && \cdot \biggr [{\mathcal S}_{Qq+}(\xi_1,\xi_2)  T_{q+} (y_1,0)   + {\mathcal S}_{Qq-}(\xi_1,\xi_2) T_{q-}(y_1,0)  \biggr ],
\nonumber\\
   {\mathcal S}_{Qq+} (\xi_1,\xi_2) &=& \frac{1}{N_c^2-1} \left ( \xi_1^2 +\xi_1\xi_2 -2\xi_1-\xi_2 +1 \right ),
\nonumber\\
      {\mathcal S}_{Qq-} (\xi_1,\xi_2) &=& \frac{1}{N_c^2-1} \left ( (\xi_1 +\xi_2)^2 +4 (1-\xi_1-\xi_2)  \right ).
\label{Fac-SFP-P1}
\end{eqnarray}

\par
\begin{figure}[hbt]
\begin{center}
\includegraphics[width=14cm]{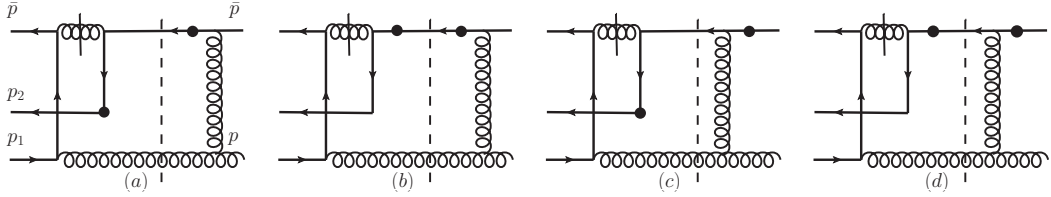}
\end{center}
\caption{The diagrams in the gauge $n\cdot G=0$ for the amplitude
$\bar q + (q + \bar q) \to \gamma^*  + X \to \bar q + G $ at one-loop for possible SFP contributions.  }
\label{SFP-P2}
\end{figure}
\par
\begin{figure}[hbt]
\begin{center}
\includegraphics[width=8cm]{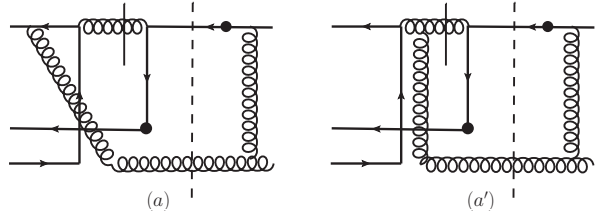}
\end{center}
\caption{The diagrams obtained from Fig.\ref{SFP-P2}a by changing the attachment of the collinear gluon.  }
\label{SFP-P2F}
\end{figure}

We turn to the $q\bar q$-contributions. The contributions are given
by diagrams in Fig.\ref{SFP-P2} in the light-cone gauge.
We need to find the collinear divergences related to the collinear gluon
crossing the cut in these diagrams. But, a direct calculation of the collinear divergences in  these diagrams
will give wrong results. This is the exception mentioned in Sect.5.1 before Eq.(\ref{WFig6}).
We will explain this with Fig.\ref{SFP-P2}a as an example.
In this diagram, the collinear divergence appears when the gluon attached to the initial quark is collinear
to the $+$-direction.
Instead of attaching the collinear gluon to the initial quark, it can also attached to other places.
There are two examples given by the diagrams in Fig.\ref{SFP-P2F}.

\par
\par
As discussed in Sect.5.1., one may expect that these two diagrams in Fig.\ref{SFP-P2F} do not have
the discussed collinear divergence in the gauge $n\cdot G=0$. Because of the structure of the color factor, Fig.\ref{SFP-P2F}a$'$ is always zero.
But, through an explicit calculation one finds that Fig.\ref{SFP-P2F}a also contains the collinear divergence.
Similarly to Fig.\ref{SFP-P2}a, we can obtain
the corresponding diagram Fig.\ref{SFP-P2F}b, Fig.\ref{SFP-P2F}c and Fig.\ref{SFP-P2F}d
from Fig.\ref{SFP-P2}b, Fig.\ref{SFP-P2}c and Fig.\ref{SFP-P2}d,
respectively. These diagrams are not drawn in Fig.\ref{SFP-P2F}. They also contain
collinear divergences.
If the divergences survive in the end results, it implies that
the factorization is broken. This needs to carefully be examined.

\par
We use $k$ to denote the momentum carried by the gluon crossing the broken line. If the gluon is collinear,
$k$ has the patten:
\begin{equation}
  k^\mu \sim (1, \lambda^2, \lambda,\lambda ), \ \ \ \  \lambda \ll 1.
\label{colk}
\end{equation}
We use $k_g$ to denote the momentum carried by the gluon propagator with the short bar. The propagator
has three terms in the light-cone gauge:
\begin{equation}
 \pi \delta (k_g^2) \left [ -g^{\mu\nu} + \frac{n^\nu k_g^\mu}{n\cdot k_g} + \frac{n^\mu k_g^\nu}{n\cdot k_g} \right ].
\label{ngp}
\end{equation}
In the above $\mu$ is the index contracted with that in the vertex left to the short bar, and $\nu$ is contracted
with that in the vertex right to the short bar.
The first term will not give collinear divergence in Fig.\ref{SFP-P2F}a. But, the second and third
term will give collinear divergences with the collinear power-counting, because the denominator
of the terms is at order of $\lambda^2$, i.e., $n\cdot k_q \sim \lambda^2$ derived from the on-shell condition
$\delta(k_q^2)$ with Eg.(\ref{colk}).
\par
The propagator in Eq.(\ref{ngp}) also appear in Fig.\ref{SFP-P2}a. The second term gives no contribution
because of $\bar v(\bar p) n\cdot \gamma =0$. The contributions from the first- and third term contain
the collinear divergences. It is easy to show that the divergence from the third term is canceled
by that from the third term in Fig.\ref{SFP-P2F}a. This also happens for other diagrams in Fig.\ref{SFP-P2}
in a similar way. Through explicit calculation we find that the divergence introduced by the second term in Fig.\ref{SFP-P2F}a and
Fig.\ref{SFP-P2F}b are canceled by that in Fig.\ref{SFP-P2F}c and
Fig.\ref{SFP-P2F}d, respectively. Therefore, only the collinear divergences in Fig.\ref{SFP-P2}
introduced by the first term
in Eq.(\ref{ngp}) survive at the end, if we include all diagrams from Fig.\ref{SFP-P2} and
Fig.\ref{SFP-P2F} in the gauge $n\cdot G=0$.  The diagrams in the light-cone gauge
by changing the attachment of the collinear gluon
in the right part of diagrams in Fig.\ref{SFP-P2} do not contain collinear divergences.
This has the implication for using the diagram expansion in the light-cone gauge, where
one will have the uncanceled divergences from the cut gluon-propagator. With the method in Feynman gauge
one will not have such divergences.

\par
From the above discussion the correct result is to obtain  by taking only the first
term in Eq.(\ref{ngp}) to calculate the diagrams in Fig.\ref{SFP-P2}, or  by taking all in Eq.(\ref{ngp})  to calculate all diagrams
inFig.\ref{SFP-P2} and
Fig.\ref{SFP-P2F}. We obtain:
\begin{eqnarray}
{\mathcal W}_T\biggr\vert_{Fig.\ref{SFP-P2}}       &=&  -\frac{e_q^2 g_s\alpha_s^2}{8\pi^2} \frac{N_c^2-1}{ N_c}
     \left ( -\frac{2}{\epsilon_c} \right )
   \frac{\sqrt{2 x_0\bar x_0}}{x_0\bar x_0 q^2_\perp}\delta ( s(\bar x_0 -x)(1-y) -q^2_\perp)
\nonumber\\
     &&  \left [ \frac{(1-y)\bar x_0 -x}{\bar x_0^2}(\bar x_0 -x)  \left ( {\mathcal C}_+^{q\bar q}
     - {\mathcal C}_-^{q\bar q} \right )
        +(x+\bar x_0 y -2\bar x_0)^2  \left ( {\mathcal C}_+^{q\bar q}
     + {\mathcal C}_-^{q\bar q} \right )  \right ].
\end{eqnarray}
With  the results of relevant twist-3 matrix element in Eq.(\ref{SQP-T3P2})  one can
derive the following factorized form:
\begin{eqnarray}
{\mathcal W}_T\biggr\vert_{Fig.\ref{SFP-P2}} &=& \frac{ e_q^2 \alpha_s}{\pi^2 N_c q^2_\perp}
\int_{x}^1 \frac{d y_1}{y_1} \int_y^1 \frac{dy_2}{y_2} \bar q(y_2) \delta (\hat s (1-\xi_1) (1-\xi_2) -q^2_\perp)
\nonumber\\
   && \cdot  \biggr [ {\mathcal S}_{Qq\bar q+} (\xi_1,\xi_2)  T_{q+} (-y_1,0) +{\mathcal S}_{Qq\bar q-} (\xi_1,\xi_2) T_{q-} (-y_1,0) \biggr  ],
\nonumber\\
     {\mathcal S}_{Qq\bar q+} (\xi_1,\xi_2) &=& -\frac{1-\xi_1}{ 2 N_c} ( 1-\xi_1-\xi_2),
     \ \ \
     {\mathcal S}_{Qq\bar q-} (\xi_1,\xi_2) = -\frac{(2-\xi_1-\xi_2)^2}{2 N_c}.
\label{Fac-SFP-P2}
\end{eqnarray}

\par
For the $gg$-contributions there are also a SQP contribution, where one can obtain $T{q\pm}(x,0)$  from
the $gg$-contributions at one-loop. The SQP contribution in ${\mathcal W}_T$ is obtained
by replacing $h_B$ with a gluon at one-loop. This contribution is in fact contained
in the factorized from in Eq.(\ref{Fac-SFP-P1}). This is similar to the case in $q\bar q$-contributions
for the SGP-contributions with Fig.\ref{SGP-P3} discussed in Sect.5.1.

\par
Combining all flavors we obtain then the factorized SQP contributions
as:
\begin{eqnarray}
{\mathcal W}_T\biggr\vert_{SQP}  &=& \frac{\alpha_s }{\pi^2 N_c  q_\perp^2}
 \int_x^1 \frac{dy_1}{y_1} \int_y^1\frac{dy_2}{y_2}
  \delta(\hat s (1-\xi_1)(1-\xi_2)-q^2_\perp)
\nonumber\\
   && \cdot \biggr [ {\mathcal S}_{Qq+}(\xi_1,\xi_2) \sum_{[q]} e_q^2 f_g(y_2)  T_{q+} (y_1,0)   + {\mathcal S}_{Qq-}(\xi_1,\xi_2) \sum_{[q]} e_q^2 f_g (y_2) T_{q-}(y_1,0)
\nonumber\\
  &&  + {\mathcal S}_{Qq\bar q+} (\xi_1,\xi_2) \sum_{[q]} e_q^2 f_{\bar q} (y_2)  T_{q+} (-y_1,0) +{\mathcal S}_{Qq\bar q-} (\xi_1,\xi_2) \sum_{[q]} f_{\bar q} (y_2)  T_{q-} (-y_1,0)  \biggr ].
\end{eqnarray}
In comparison with the existing results in \cite{KKnew}
derived with the method of diagram expansion our results of SQP contributions are different.
The difference
is of an overall factor of $-2$. We note that the SQP contribution is proportional to $q_\perp^{-2}$ in the limit
$q^2_\perp \to 0$. Hence, it is not a leading contribution in the limit $q^2_\perp /Q^2 \ll 1$.

\par\vskip20pt
\noindent
{\bf 7. Summary}
\par\vskip10pt
\par\vskip10pt
\par
We have studied the collinear factorization of SSA in Drell-Yan processes.
To derive all perturbative coefficient functions at leading order of $\alpha_s$ 
in the factorization, we have studied the scattering with
multi-parton states, in which the helicity of the states are flipped.
SSA in such a scattering is nonzero.
This is in contrast to the scattering with a transversely polarized single quark.
In this case SSA is always zero because of the helicity conservation
of QCD for massless quarks.
\par
We have calculated SSA in the multi-parton scattering processes and the
relevant twist-3 matrix elements of multi-parton states. By using the results
from our calculation SSA has been factorized as convolutions of twist-3 matrix elements
of the polarized hadron, parton distribution functions of the unpolarized hadron and perturbative
coefficient functions.  All perturbative coefficient functions
of these contributions are derived here at the leading order of $\alpha_s$.
In the factorization there are HP-, SGP- and SFP-contributions.
From our results, we find that SSA at tree-level
is factorized as the HP contributions.
But the SGP- and SFP- contributions are from a class of one-loop contributions to SSA.
These one-loop contributions contain collinear divergences and they can only be factorized
with the soft-pole twist-3 matrix elements in which one of the active patrons carries zero momentum.
These soft-pole twist-3 matrix elements are zero at tree-level but nonzero at one-loop.
This results in that the perturbative coefficient functions of SGP- and SQP contributions are at
the same order as those of HP contributions. Hence, in the collinear factorization
there is a nontrivial order-mixing. Such an order-mixing does not happen in the factorization
only involving twist-2 operators.
\par
It is interesting to note that at one-loop SSA contains divergences caused by exchanges of a Glauber gluon,
as discussed in Sect.5. The divergences are factorized with the soft-gluon-pole matrix elements.
This is in contrast to the factorization of unpolarized cross-section only with twist-2 operators,
where it is well known that the divergences from exchanges of Glauber gluons are canceled\cite{G0,G1,G2}.
In the case of SSA studied here with twist-3 operators, such divergences are not canceled
and need to be factorized.
This will have some implications for the study of factorizations in the framework of soft collinear effective theories of QCD\cite{LiuMa}.
\par
Our results for the collinear factorization of SSA in Drell-Yan processes agree with those
derived with the method of diagram expansion, except the SQP contributions studied in Sect.6. 
Comparing the method of the diagram expansion,
we believe that it has advantages to use our method with multi-parton states
for analyzing factorizations of SSA and for calculating higher order corrections, because the involved
calculations are of standard scattering amplitudes. The approach we have taken here provides
another way to derive the collinear factorization of SSA in various processes. It will be useful
to solve the discrepancy between results for SSA in \cite{SSA0QT}, where the momentum of a lepton
in Drell-Yan processes is measured. It will also be useful for solving the discrepancy of evolutions of
twist-3 matrix elements derived in \cite{KQ,BMP,ZYL}. We leave these for future work.

\par\vskip20pt
\noindent
{\bf Note Added:} During the preparation of the paper  the results of 
the SGP-contributions with gluonic twist-3 matrix elements is reported 
in \cite{KY3G}. The results there agree with ours in Sect. 5.2..

\vskip 5mm\par

\par\noindent
{\bf Acknowledgments}
\par
This work is supported by National Nature
Science Foundation of P.R. China(No. 10975169,11021092). The work of H.Z. Sang is supported
by the Fundamental Research Funds for the Central Universities(WM1114025) and by National Nature
Science Foundation of P.R. China(No. 11147168).

\par\vskip30pt

\par

\end{document}